\documentclass{article}
\usepackage[english]{babel}

\usepackage[letterpaper,top=2cm,bottom=2cm,left=3cm,right=3cm,marginparwidth=1.75cm]{geometry}

\usepackage{url}
\usepackage{cite}

\usepackage{curve2e}
\usepackage{amsfonts}
\usepackage{amssymb}
\usepackage{amsmath}

\usepackage{chemformula}
\usepackage[labelformat=simple]{subcaption}
\usepackage{graphicx}
\usepackage{grffile}
\usepackage{caption}
\usepackage{subcaption}
\usepackage{bm}

\usepackage{hyperref}

\usepackage{xcolor}

\usepackage{ulem}

\usepackage{authblk}
\title{Charge-Density Waves vs. Superconductivity:\\ Some Results and Future Perspectives}

\author[1]{Giulia Venditti}
\author[2]{Sergio Caprara}
\affil[1]{SPIN-CNR Institute for Superconducting and other Innovative Materials and Devices,\\ Area della Ricerca di Tor Vergata, Via del Fosso del Cavaliere 100, 00133 Rome, Italy}
\affil[2]{ISC-CNR and Department of Physics, Sapienza University of Rome, P.le A. Moro 5, 00185 Rome, Italy}
\date{\today}                     
\begin{document}
	
\maketitle

\begin{center}
Correspondence: giulia.venditti@spin.cnr.it
\end{center}

\begin{abstract}
Increasing experimental evidence suggests the occurrence of filamentary superconductivity in different (quasi) two-dimensional physical systems. In this piece of work, we discuss the proposal that under certain circumstances, this occurrence may be related to the competition with a phase characterized by charge ordering in the form of charge-density waves.  
{We provide a brief summary of experimental evidence supporting our argument in two paradigmatic classes of materials, namely transition metal dichalcogenides and cuprates superconductors. We present a simple Ginzburg--Landau two-order-parameters model as a starting point to address the study of such competition. We finally discuss the outcomes of a more sophisticated model, already presented in the literature and encoding the presence of impurities, and how it can be further improved in order to really address the interplay
	between charge-density waves and superconductivity and the possible occurrence of filamentary superconductivity at the domain walls between different charge-ordered regions.}
\end{abstract}

\section{Introduction}
\label{sec:intro}
The physics of (quasi) two-dimensional superconductors is attracting 
increasing attention thanks to the recent advances in the synthesis 
and tailoring of two-dimensional electron systems. Among~these, 
we mention oxide interfaces~{\cite{hwang2012emergent}}, field-effect-controlled devices~{\cite{schmidt2015electronic}}, 
atomic monolayers grown by means of molecular beam epitaxy~{\cite{rajan2020morphology}}, and 
thin flakes obtained by exfoliation~{\cite{yang2023synthesis}}. Nowadays, it 
is even possible to single out and combine two-dimensional materials, such as graphene, individual layers 
of transition metal dichalcogenides and
high-critical-temperature superconducting cuprates, paving the way toward the engineering of functionalized two-dimensional 
systems~\cite{geim2013van}.

All the above systems share the common feature of being highly 
crystalline and therefore represent the ideal context to study 
intriguing phenomena~\cite{Saito2016}
, such as, e.g.,~Ising or topological 
superconductivity, bypassing or at least reducing the 
obnoxious effect of defect-induced~disorder. 

One of the noticeable properties of many two-dimensional systems is 
the occurrence of a low-temperature residual metallic state, 
which has been proposed as a new quantum state of matter, i.e.,~the so-called {``quantum metal''}~\cite{tsen2016nature} or {``anomalous metal'' \cite{kapitulnik2019colloquium}}, which escapes
the 
customary fate of two-dimensional electron systems, which are 
either Anderson insulating or superconducting at low temperatures.
{The residual metallic phase is reached upon lowering the temperature from the high-temperature metallic state and passing through an intermediate regime where the resistance is significantly  
	suppressed, which is seemingly due to incipient superconductivity. 
	Superconductivity is, however, unable to fully develop, e.g.,~due to the presence of a magnetic field or~to other mechanisms at play.}
{It is worth noting that this behavior has been observed also in highly crystalline two-dimensional systems, such as transition 
	metal dichalcogenides or ultrathin ZrNCl films~\cite{Dezi2018}.}
Some features of the resistivity vs. temperature curves $\rho(T)$
are characteristic of this peculiar state. Indeed, even if the conditions for a full development of superconductivity are met, e.g.,~because the external magnetic field is not strong enough, and~the zero-resistance state is finally
reached, the~metal-to-superconductor transition is so broad that its width cannot be attributed to standard fluctuating 
phenomena, such as those described by the Aslamazov--Larkin or 
Halperin--Nelson theory, with~any reasonable choice of the 
phenomenological parameters~\cite{caprara2011effective}. 
Moreover, the~$\rho(T)$ curves display
a rather pronounced tail on their low-temperature side. The~presence
of tails in the $\rho(T)$ curves is customarily associated to
vortex-driven dissipation~\cite{tinkham2004introduction}; however, those tails are observed also in the absence of a 
magnetic field, for~instance when the low-temperature 
superconducting phase is weakened by a (gate-controlled) reduction
of the carrier density. This very fact witnesses that the 
customary mechanism of vortex-driven dissipation cannot be 
held responsible for the pronounced tails in the $\rho(T)$ curves.

An effective and successful interpretation for these anomalous features 
is offered by the comparison with the case of the 
two-dimensional  electron gas formed at oxide interfaces 
such as, e.g.,~LaAlO$_3$/SrTiO$_3$, where the $\rho(T)$ curves
exhibit strikingly similar behavior. 
Despite the highly crystalline structure of LaAlO$_3$/SrTiO$_3$ 
interfaces, the~occurrence of tails in their $\rho(T)$ curves
has been successfully interpreted in terms of nanoscale electron 
inhomogeneity~\cite{biscaras2013multiple, bucheli2013metal, caprara2013multiband, prawiroatmodjo2016evidence}. 
Further support to the statement that oxide 
interfaces and other two-dimensional electron systems are 
(intrinsically) inhomogeneous comes from the observation of a 
low-temperature Griffiths state when the metal-to-superconductor 
transition {can be} driven by a magnetic field~\cite{shen2016observation, Saito2018}. 
The claim of inhomogeneity might apparently clash with the highly crystalline structure of these systems. 
It is the good mobility of the charges that is at seeming odds with the transport observations, weakening customary scattering arguments that can be raised in metals at low temperatures (crystal defects and impurities).
Instead, inhomogeneity may have an intrinsic origin~\cite{caprara2012intrinsic, scopigno2016phase} due to some mechanism
that endows the electrons with a tendency to segregate, at~least on
a nanoscopic scale, resulting in a landscape of tiny but significant
electron density modulations that are apt to locally enhance or 
suppress superconductivity~\cite{caprara2015interplay}.

{Strong inhomogeneities of the superconducting condensate can indeed appear as a 
	filamentary cluster whose origin might arise from different (intrinsic and/or extrinsic) sources, one of which might
	be the competition with another ordered state of matter. In~this piece of work, we want
	to explore the scenario where this competing 
	state is characterized by charge ordering 
	in the form of charge-density waves. This competing mechanism seems to be relevant to
	at least two classes of (quasi) two-dimensional systems, namely
	some transition metal dichalcogenides and high-critical-temperature 
	superconducting cuprates. }

{We point out that hints of filamentary superconductivity have been reported in
	other classes of materials, among~which we find iron-based superconductors~\cite{li2021pressure, xiao2012evidence, xiao2012filamentary, gofryk2014local}, whose phase diagram is very similar to that of cuprates, with~an antiferromagnetic phase at low doping and a superconducting dome at higher hole doping, a~pseudogap region and a stripe phase. Nonetheless, no evidence of charge ordering was found so far in iron-based materials, although~spin-density waves could be a good candidate to consider in order to explore a similar scenario also in these systems. As~a matter of fact, the~stripe phase in iron-based superconductors has already been linked to the presence of spin-density waves~\cite{machida1989magnetism, kato1990soliton,  carlson2004spin, carlson2008low}. }

The structure of the present paper is the
following. In~Section~\ref{TMDs}, we discuss evidence for 
the competition between superconductivity and charge-density 
waves in transition metal dichalcogenides, while in 
Section~\ref{cuprates} we discuss the case of 
high-critical-temperature superconducting cuprates. In~
Section~\ref{sec:GL-model}, we introduce a simple Ginzburg--Landau
model to describe the competition of the two phases. In~
Section~\ref{sec:competition}, we discuss in some more detail the
physics of this competition. Our concluding remarks, as~well as some directions to improve our understanding
of the competition between superconductivity and charge-density waves, are found in Section~\ref{sec:conclusions}.

\section{Experimental evidence for the competition between	Superconductivity and Charge-Density Waves in Some Selected~Systems}
As we have suggested above, there are at least
two classes of (quasi) two-dimensional materials 
where the competition between superconductivity and 
charge-density waves may play a relevant role, namely
some transition metal dichalcogenides and 
high-critical-temperature superconducting cuprates. Experimental evidence for such a competition to occur in these two paradigmatic systems
is discussed below in~Sections~\ref{TMDs}
and \ref{cuprates}, respectively.

\subsection{Transition Metal~Dichalcogenides}
\label{TMDs}

\begin{figure}[t]
	\centering
	\includegraphics[width=0.8\linewidth]{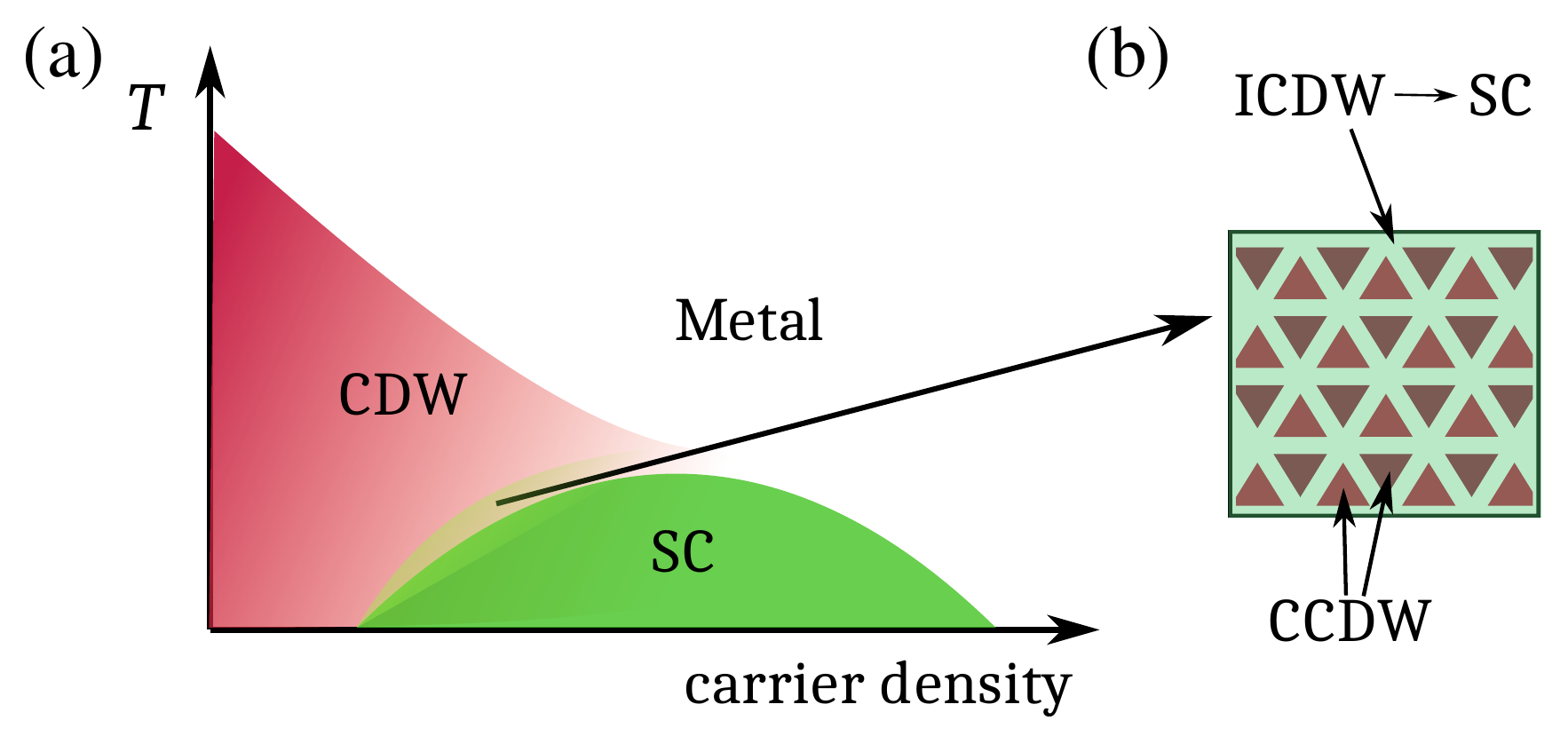}
	\caption{{(\textbf{a}) {Cartoon} 
			sketch of the phase diagram for transition metal dichalcogenides and (\textbf{b}) cartoon of the formation of a pattern of commensurate 
			charge-density-wave regions (CCDW) embedded in an incommensurate charge-density-wave (ICDW) matrix, the~latter possibly becoming superconducting. }}
	\label{fig:TMD}
\end{figure}

The observation of periodic oscillations of the magnetoresistance 
induced by the Little–Parks effect in 1T-TiSe$_2$~\cite{Li2016a} 
shows that the onset of superconductivity is directly 
related to the spatial texturing of the amplitude and phase 
of the superconducting order parameter, corresponding to a 
two-dimensional superconducting matrix. The~authors of
Ref.~\cite{Li2016a} 
infer that such a superconducting
matrix originates from a matrix of incommensurate 
charge-density-wave states embedded in the commensurate 
charge-density-wave states, and~they argue that their results 
give evidence for spatially modulated electron states to be 
fundamental to the appearance of two-dimensional superconductivity 
in 1T-TiSe$_2$ {(see Figure~\ref{fig:TMD})}. Indeed, it is clear that the peculiar features of 
the magnetoresistance mirror the spatial fluctuations of the 
superconducting (Cooper) pairing. A~reasonable explanation for 
these peculiarities is based on the Little–Parks effect, whereby 
Cooper pairs are constrained to move in loops, forcing
pairings to have a local character and to occur in well-defined regions. The~observed length scale is associated with 
Cooper pairs trapping magnetic flux quanta. The~remarkable 
occurrence of such a superconducting matrix in a single
crystal calls for a pre-existing matrix of inhomogeneous 
electron states to stabilize it. Fluctuations of an underlying 
charge or spin order parameter appear to be relevant to the 
occurrence of superconductivity in a variety of physical systems. 
The suppression of the charge-density-wave transition from 
$T_{\text CDW}=170$\,K to $T_{\text CDW}=40$\,K, 
alongside with the appearance of superconductivity and marked
non-Fermi-liquid behavior of the metallic phase, strongly 
suggest that the charge-density waves play a role in 1T-TiSe$_2$.
The authors of Ref.~\cite{Li2016a}
speculate that domain 
walls form a periodic matrix in which commensurate 
charge-density-wave domains with constant phases are embedded 
in an incommensurate charge-density-wave~matrix. 

{Analogously, Little--Parks effect signatures have been reported also in Li$_x$-TiSe$_2$~\cite{liao2021coexistence}, where magnetoresistance oscillations were observed in the superconducting phase along with an ``anomalous metal'' behavior.
	The transport characterization of electric field-controlled lithium intercalation in TiSe$_2$ provided evidence for the coexistence of commensurate and incommensurate charge-density waves, which is supported by the correlation between the charge-density wave and the magnetoresistance oscillations. 
	Thus, also in this case, the~authors speculate a periodic structure of commensurate charge-density waves encoded in an incommensurate 
	charge-density-wave network, whereby the latter 
	becomes superconducting at low temperatures. 
	The average size $L$ of the commensurate regions was thus extrapolated in Ref.~\cite{liao2021coexistence}, as~in Ref.~\cite{Li2016a}, by~comparing the magnetic field oscillations $\Delta B$ with the flux quantum, providing an opposite temperature and doping dependence of $L$ with respect to the case of ion-gated TiSe$_2$ presented in Ref.~\cite{Li2016a}. While in fact, in Ref.~\cite{Li2016a}, $L$ quickly decreases with the carrier density, an~increasing trend is observed in Ref.~\cite{liao2021coexistence}.
	The other important difference between those two materials is their phase diagram. In~\cite{liao2021coexistence}, $T_\text{CDW}$ is in fact constant as a function of the carrier density, whereas instead, the intensity of the charge-density-wave response $I_\text{CDW}$ is suppressed with lithium intercalation.} 

{The same trend of a constant $T_\text{CDW}$ and a decreasing $I_\text{CDW}$ is found in electric field-driven hydrogen intercalated TiSe$_2$, which is probed through transport and magnetotransport measurements, X-ray diffraction, Raman spectroscopy and nuclear magnetic resonance~\cite{piatti2023superconductivity}. 
	As in the previous case, the~insensitivity of $T_\text{CDW}$ to increasing doping can be interpreted as a lack of full in-plane percolation of the dopants occurring on length scales smaller than $\sim$10 ${mu}$m.
	On the other side, the~charge-density-wave order might still appear in H-rich regions; two possible explanations could be either that the dopants suppresses long-range order and the charge-density waves survive only on short-range scales, or, similarly to the case of Cu$_x$-TiSe$_2$ \cite{spera2019energy}, that the opening of the charge-density-wave gap is shifted below the Fermi level with an increasing density of charge carriers.
	Concerning instead the superconducting phase, hydrogen-intercalated TiSe$_2$ does not show 
	the dome behavior in doping that characterizes other transition metal dichalcogenides, 
	the superconducting critical temperature $T_\mathrm{c}$ being finite and constant as the doping is varied.}

{A structure similar to the ion-gated TiSe$_2$} was observed by scanning transmission microscope 
measurements of the closely related 1T-TaS$_2$, in~which the 
incommensurate charge-density-wave state is found even at ambient 
conditions~\cite{burk1991charge}. Self-organization results 
from the repulsive interactions between domain walls, which are associated 
with higher-order terms in the free energy~\cite{mcmillan1975time}. 
Therefore, the~incommensurate charge-density-wave state will 
form a matrix, fragmenting the commensurate charge-density-wave 
state into domains with fixed area, as~needed for the occurrence
of the Little--Parks effect. As~shown in 
Ref.~\cite{mcmillan1975time}, incommensurate 
charge-density-wave dynamic phase fluctuations, i.e.,~the 
phonon modes of the incommensurate charge-density wave, can 
occur within the domain walls. 
It is plausible that these incommensurate charge-density-wave phonons may trigger superconducting pairing and localize the resulting Cooper pairs 
in one-dimensional regions of the two-dimensional system. 
Another intriguing result of Ref.~\cite{mcmillan1975time}
is that the point-contact conductance spectra measured at 
each carrier density is characterized by the presence of a 
zero-bias conductance peak in the superconducting state. 
Zero-bias conductance peaks are observed in a variety of 
seemingly unconventional superconductors and are interpreted
as the consequence of Andreev reflection promoted by a Cooper 
pairing potential with an internal phase shift of 
the superconductivity order parameter~\cite{kashiwaya2000tunneling}. 

The coexistence of commensurate and incommensurate charge-density waves was first observed by X-ray measurements on 
TiSe$_2$ at pressures close to those at which superconductivity 
was expected to occur~\cite{Joe2014}
. Incommensurate charge-density-wave domain walls with a periodicity along the $c$ axis of approximately
300\,nm were observed, which is similar to the length scale determined in 
Ref.~\cite{Li2016a}.

X-ray diffraction measurements on Cu$_x$-TiSe$_2$ \cite{Kogar2017} 
highlighted an incommensurate charge-density wave, occurring at an intercalant concentration 
which coincided with the onset of superconductivity. This 
result seems to agree with an increasing number of experiments 
pointing to the relevance of incommensurate charge-density waves
for the occurrence of superconductivity in transition metal
dichalcogenides~\cite{Joe2014, liu2013superconductivity, Sipos2008b, mutka1983superconductivity}
, although~the role of crystallographic disorder could not be ruled out. In~addition, the~authors of Ref.~\cite{Kogar2017} 
showed that the charge-density waves do not terminate near (or 
inside) the superconducting dome, and~rather, they survive up to 
an intercalant concentration much larger than previously~thought.

It is worth noting that in 1T-TaS$_2$, the~commensurate 
charge-density wave is destabilized by pressure or Li
ion intercalation, whereas the incommensurate 
charge-density wave survives and coexists with 
superconductivity~\cite{Sipos2008b}. Likewise, in~1T-TaS$_{2-x}$Se$_x$, 
superconductivity is sandwiched between two regions of
commensurate charge-density waves as $x$ is varied, and~
again, it only coexists with incommensurate charge-density waves~\cite{liu2013superconductivity}
. In~the case of
2H-TaSe$_2$, the~superconducting transition temperature
can be raised to 2\,K by irradiating the sample, 
thereby disrupting the commensuration of the charge-density
wave and introducing disorder~\cite{mutka1983superconductivity}. 
Interestingly, the~incommensuration of the charge-density
wave can be distinguished only at about the same 
intercalation content at which superconductivity sets in. This 
is similar to what is observed in TiSe$_2$ under pressure, 
where incommensurate fluctuations were highlighted above 
the superconducting dome~\cite{Joe2014}. 

Likely, the~peculiarities of the phase diagram of {Cu$_x$-TiSe$_2$}
are the consequence of Cu intercalants electron doping the Ti-$3d$
conduction band~\cite{qian2007emergence}, 
thereby suppressing excitonic correlations. Many 
studies led to the conclusion that both electron--phonon coupling 
and electron--hole coupling drive the charge-density-wave
transition in pure 1T-TiSe$_2$~\mbox{\cite{di1976electronic, kidd2002electron, cercellier2007evidence, hellmann2012time, rohwer2011collapse, Rossnagel2011, weber2011electron, Porer2014,  van2010alternative, van2010exciton}}. 
Electron doping selectively weakens the 
excitonic contribution to the charge-density wave by~shifting 
the chemical potential into the conduction band, thereby enhancing 
screening effects, while leaving the electron--phonon interaction
substantially unaffected. This interpretation provides a natural
framework to interpret the peculiarities of the phase diagram 
of {Cu$_x$-TiSe$_2$}: both excitonic and electron--phonon 
interactions drive the charge-density-wave transition in the 
low-intercalation region of the phase diagram, while only 
electron--phonon coupling plays a role in the
high-intercalation~region. 

Scanning tunneling microscopy measurements~\cite{spera2019energy},
meant to investigate the interplay between charge-density waves 
and superconductivity in 1T-Cu$_x$TiSe$_2$, confirmed that 
the implication of Cu atoms in the observed alterations of 
charge-density waves is challenging, because~Cu atoms and 
charge-density waves cannot be simultaneously probed. Indeed, 
Cu atoms are only resolved at negative bias voltages, less than
$-800$\,mV~\cite{novello2017stripe, yan2017influence},
while charge-density-wave contrast 
is achieved at lower bias voltages within~a few hundred meV of 
the charge-density-wave gap~\cite{novello2015scanning}. 
To obtain the alignment of images taken at such different biases, 
fingerprints of atomic defects visible at all biases, in~particular intercalated Ti, were used~\cite{novello2015scanning, hildebrand2014doping}. 
An interesting feature of the  
charge-density waves in Cu-intercalated TiSe$_2$ is the presence 
of an inhomogeneous electron background. Such an inhomogeneity 
is directly related to intercalated Cu atoms, which tend to cluster.
Seemingly, Cu intercalation affects the long-range $2\times 2$ 
commensurate charge-density wave observed in the $ab$ plane 
of pristine crystals in two different ways: it induces a 
sizable energy-dependent patchwork of
charge-density-wave regions while promoting the
formation of $\pi$-phase shift domain walls. The~contrast 
inversion expected for a standard electron--hole symmetric 
charge-density wave~\cite{dai2014microscopic} 
is~missing.

Finally, the~charge-density-wave pattern sheds new light on the properties of ion-liquid-gated TiSe$_2$ films~\cite{Li2016a}, 
resulting in a spatially inhomogeneous carrier distribution~\mbox{\cite{costanzo2016gate, petach2017disorder, Dezi2018}};
the associated nonuniform potential
landscape is expected to promote energy-dependent 
charge-density-wave patches, similar
to those observed in Ref.~\cite{spera2019energy}.

The observed charge-density-wave pattern provides evidence 
that the charge-density-wave gap in 1T-Cu$_x$TiSe$_2$ opens below 
the Fermi level and~moves to higher binding energies with 
increasing Cu content. Remarkably, the~charge-density wave probed by 
scanning tunneling microscopy survives for Cu
doping deep inside the superconducting dome, pointing at a
possible coexistence of the two~phases.

\subsection{High-Critical Temperature Superconducting~Cuprates}
\label{cuprates}

\begin{figure}[t]
	\centering
	\includegraphics[width=0.8\linewidth]{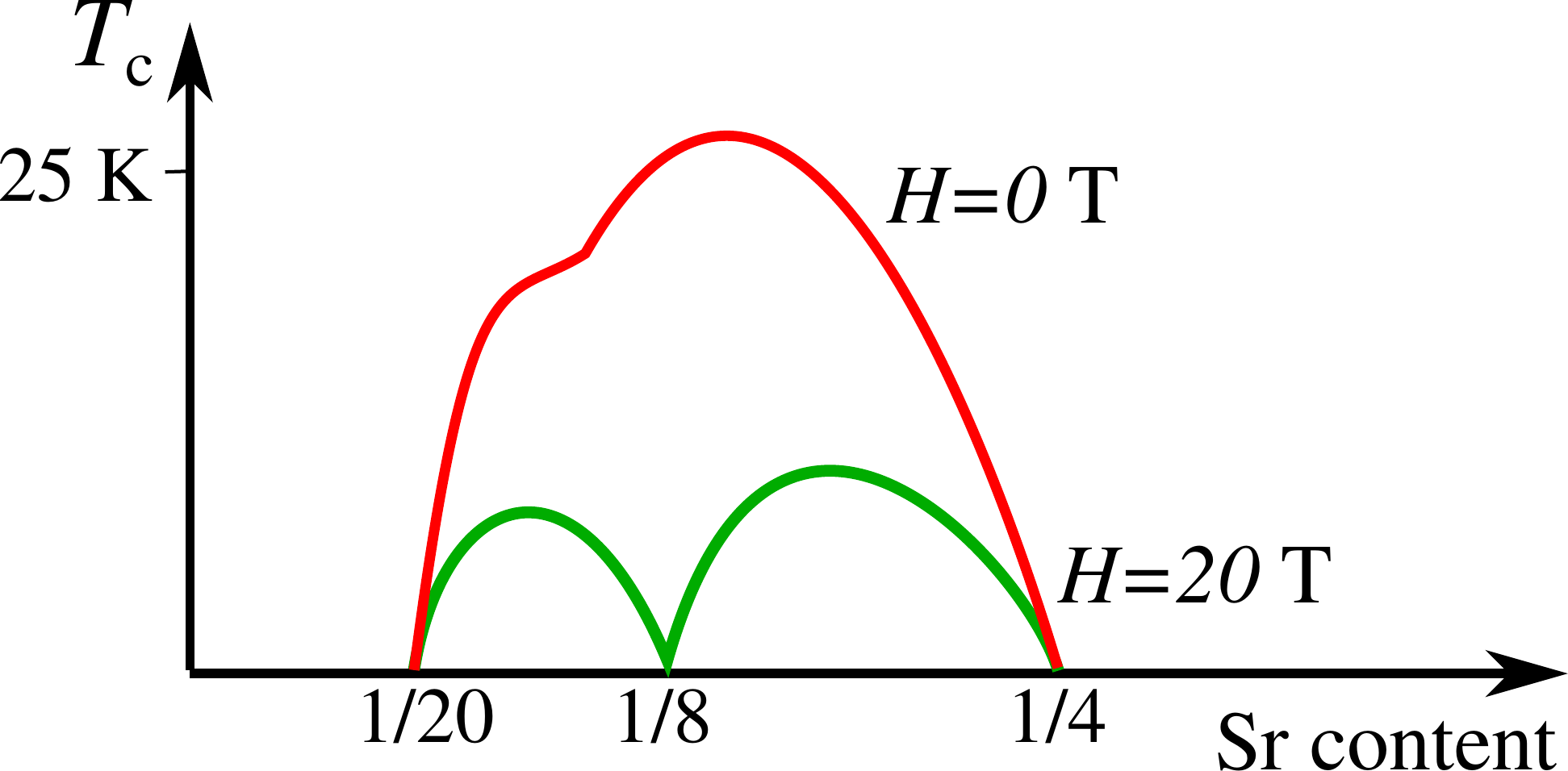}
	\caption{Superconducting critical temperature $T_c$ (defined as 
		the  occurrence of a zero-resistance state) of LSCO as a function of doping (Sr content) in~the absence of magnetic field 
		(red curve, labeled with $H=0$\,T) and for a magnetic field $H=20$\,T (green curve). The~presence of two distinct domes at high magnetic field was also observed in YBCO in {Ref}.~\cite{grissonnanche2014direct}. The~sketchy figure is adapted from {Ref}.~\cite{caprara2020doping}, where $T_c$ curves at intermediate values of the
		magnetic field and other details can be found.}   
	\label{fig:LSCO_phasediag}
\end{figure}

The phase diagram of high-critical temperature superconducting 
cuprates exhibits a variety of competing phases. Hereafter, we
discuss the more common hole-doped cuprates. The~undoped parent 
compound is an antiferromagnetic Mott insulator, but~antiferromagnetism is rapidly suppressed with increasing doping, 
the N\'eel temperature vanishing for doping $p\geq 0.02$. 
The superconducting critical temperature $T_c$ increases and 
then decreases upon doping, giving rise to a characteristic 
dome-shaped curve $T_c(p)$ in the temperature vs. doping phase 
diagram (see Figure~\ref{fig:LSCO_phasediag}). The~optimal
doping corresponds to the highest $T_c$; samples with lower 
(higher) doping are said to be underdoped (overdoped). In~the 
underdoped region of the phase diagram, the~electron
density of states at the Fermi energy is partially suppressed 
below the so-called pseudogap crossover temperature 
$T^*$ \cite{Timusk1999b}, with~$T^*(p)$ decreasing with increasing 
doping and reaching $T_c$ around optimal doping. 
Polarized neutron scattering experiments suggest a time-reversal 
and inversion symmetry breaking below $T^*$ in YBa$_2$Cu$_3$O$_{6+\delta}$ (YBCO) \cite{Fauque2006, mangin-thro2017ab, mook2008observation, baledent2011evidence}  that might be mirrored in the symmetry of the superconducting state~\cite{leridon2007josephson}. 
This symmetry breaking was subsequently observed in 
La$_{2-x}$Sr$_x$CuO$_4$ (LSCO) \cite{baledent2010twodimensional}, 
Bi$_2$Sr$_2$CaCuO$_{8+\delta}$ (BSCCO) 
\cite{dealmeida2012evidence, mangin-thro2014characterization}, and~HgBa$_2$CuO$_{4+\delta}$ (HBCO)~\cite{li2011magnetic,tang2018orientation}.

Under a strong magnetic field, the~very topology of the Fermi 
surface seems to be strongly affected in the underdoped region 
of the phase diagram, where magnetotransport studies~\cite{doiron2007quantum, doiron2015evidence, sebastian2014normal, ramshaw2015quasiparticle} 
suggest that a reconstruction of the Fermi surface takes place in 
YBCO in the doping range $0.08\lesssim p\lesssim 0.16$.  
More recent work on YBCO also suggests that a sudden change in 
the carrier density occurs around the endpoint of the line
$T^*(p)$, at~$p\approx 0.19$~\cite{badoux2016change}, which is in~agreement 
with an early scenario based on optical conductivity computations~\cite{lorenzana1993optical}, suggesting a restructuring of the Fermi surface 
with a sizable reduction of its volume at low doping.
Under the the same conditions, a~static charge ordering is 
observed by nuclear magnetic resonance~\cite{wu2012charge} and by X-ray scattering~\cite{Gerber2015}, setting in at a temperature lower than 
$T_c(p)$. Above~$T_c$, resonant X-ray scattering experiments~\cite{arpaia2019dynamical, keimer2015quantum, comin2016resonant, Miao2017, peng2017, miao2019formation} revealed the presence of dynamical charge-density waves, and nuclear magnetic resonance experiments~\cite{wu2015incipient} indicate that
static short-range charge-density waves may be pinned by the 
existing disorder, even at low magnetic field.
This rich phenomenology naturally suggests the interplay of 
various physical mechanisms and produced a wealth of 
different theoretical proposals. 
For instance, the~pseudogapped state observed below $T^*(p)$ might 
be due to an exotic metallic state resulting from strong 
electron--electron correlations in the proximity of the 
Mott insulating phase~\cite{sachdev2016novel}, or~it might 
be related to a missed quantum critical point, hidden underneath 
the superconducting dome and associated to some ordered state, 
such as circulating currents~\cite{Varma1999, varma2006theory} or 
charge order~\cite{Zaanen1989a,Castellani1995, lorenzana2002, castellani1996non, kivelson2003howto, caprara2017dynamical}.

Some authors suggest that antiferromagnetic fluctuations can play 
a role~\cite{abanov2003quantum}, despite the fact that the endpoint of the
antiferromagnetic critical line is located in the underdoped region 
outside the superconducting~dome. 

Other authors propose that the presence of Cooper pairs might 
account for the properties of the pseudogapped state. However,
the existence of preformed Cooper pairs below $T^*$ has been questioned~\cite{leridon2007paraconductivity,bergeal2008pairing}, and~
paraconductivity measurements in the pseudogapped state of LSCO~\cite{leridon2007paraconductivity, Caprara2005, caprara2009paraconductivity} show that 
customary Cooper pair fluctuations occur over a wide 
temperature range above $T_c$.

In Ref.~\cite{caprara2020doping}, a~thorough survey of the 
resistance measurement under a strong magnetic field in LSCO thin 
films was undertaken. The~ samples covered a wide doping range,
from heavily underdoped to heavily overdoped 
($0.045\le x\le 0.27$). The~strong magnetic fields was used 
to tilt the balance between different phases across the whole 
doping range. The~resulting phase diagram was in excellent 
agreement with a scenario in which, at~low doping, disorder drives 
filamentary superconductivity inside an otherwise charge-ordered phase~\cite{leridon2020protected}, which was conjectured to be a 
charge-density-wave phase, despite the fact that 
magnetotransport experiments cannot unambiguously identify the 
order parameter, and~other scenarios involving different ordered 
states~\cite{laliberte2016origin,sachdev2016novel} could not be ruled~out.


We remark that what appears as a single dome at zero and low $H$, for~large
enough $H$ splits into two domes that are centered, respectively, around  
Sr contents $x \approx 0.09$ and $x \approx 0.19$ (with a maximum at 
$x\approx 0.16$), similarly to what was 
previously observed in YBCO~\cite{grissonnanche2014direct} and also 
inferred in Ref.~\cite{campi2015inhomogeneity}. The~separation between the 
two domes is located at $x= \frac{1}{8}$. 
Two possible scenarios arise~\cite{caprara2020doping}: 
(i) commensuration effects may favor 
charge-density waves at $x=\frac{1}{8}$, producing a dip in the 
$T_c(x)$ dome, and~a stronger magnetic field weakens superconductivity
and deepens the dip, until~the superconducting dome is split in two; (ii) increasing 
the magnetic field suppresses superconductivity, thereby favoring the competing 
charge-density-wave state, except~near the endpoints $x_1$ and $x_2$ of the 
charge-density-wave dome hidden underneath the superconducting dome, where the 
strong charge fluctuations around the quantum critical points would enhance 
pairing~\cite{perali1996dwave}, thereby 
strengthening superconductivity around $x_1 \approx 0.09$ and $x_2 \approx 0.16$, 
similarly to what happens, e.g.,~in many heavy-fermion metals, where 
superconductivity arises near a quantum critical point~\cite{pfleiderer2009superconducting}.

Other hints for the interplay between superconductivity and another ordered state in cuprates come from the observation of a
two-stage transition in LSCO~\cite{shi2014two}, 
spin susceptibility measurements in YBCO~\cite{zhou2017spin} and
specific heat measurements in YBCO~\cite{kavcmarvcik2018unusual}, all suggesting charge-density waves to be the competing~phase.

{It must be pointed out that cuprates are very complicated 
	systems, the~various families sharing common aspects but~also significant differences, 
	and a lot of confusion arises about their properties, especially when the discussion is about ordered states that may compete with superconductivity. Our work falls within a line of research in which cuprates 
	of all families are on the verge of choosing between superconductivity and (some form of) charge order, and~superconductivity finally takes over. Remnants of the tendency to charge order survive in the form of dynamical charge-density waves, as~detected by RIXS~\cite{arpaia2021charge}, unless~something is done, purposely or accidentally, to~stabilize them by means of a non-thermal parameter, such as, e.g.,~Nd-codoping, or~a magnetic field, that suppresses superconductivity, uncovering static charge order underneath the superconducting dome. 
	We anticipate here that an intriguing scenario arises since the two competing phases must have very similar (free) energies, the~competition being tilted under the action of rather weak non-thermal disturbances, so the system is not far (in parameter space) from a situation when the two phases are degenerate (see Section~\ref{sec:GL-model}). 
	Likely, also structural changes may tilt the balance between superconductivity and the charge-density wave as a side effect~\cite{chen1991low, tidey2022pronounced}.
	If the two phases coexist, one as the stable phase and the~other as a metastable phase, 
	disorder can promote the formation of domains with different realizations of charge order~\cite{campi2015inhomogeneity}, and filamentary superconductivity may occur as a topologically protected parasitic phase at the domain walls between two regions with different realizations of charge order (see Section~\ref{sec:competition}). 
}

\section{The Ginzburg--Landau Model for the Competition of Two Phases}
\label{sec:GL-model}

In this section, aiming at investigating the possible outcomes
within a scenario of competing superconductivity and charge-density 
waves, we adopt a Ginzburg--Landau approach. The~free 
energy describing a system with two competing order parameters is
\begin{equation}
\label{eq:GL}
F(\Delta, \varphi)=\frac{a}{2}\Delta^2 + \frac{b}{4} \Delta^4 + \frac{c}{2} \varphi^2 + \frac{d}{4}\varphi^4 + \frac{e}{2}\Delta^2 \varphi^2
\end{equation}
where $\Delta$ ($\varphi$) corresponds to the superconducting 
(charge-density-wave) order parameter, the~last term setting 
the competition if $e>0$. {In the simplest realization, the~symmetry of the two order parameters is $Z_2$ 
	(Ising-like), but~the generalization to $U(1)$ symmetry of the superconducting order parameter is straightforward, promoting $\Delta$ 
	to a complex order parameter and interpreting $\Delta^2$ as $|\Delta|^2$ in Equation~(\ref{eq:GL}). 
	At the mean-field level, anyway, in~the absence of external fields coupled to the superconducting order parameter, one can 
	always chose the phase
	of $\Delta$ to be zero. We also point out that 
	the Ginzburg--Landau theory of a $d$-wave superconductor is essentially the same 
	as that of a $s$-wave superconductor, so our theory applies to both. Time reversal symmetry is not broken in 
	the simplest realization of our scenario, while of course, more elaborated versions can be worked out and endowed with more exotic behaviors. Our}
model was extensively studied by 
Imry~\cite{imry1975statistical} {and, more recently, by~Lee and collaborators in the Supplementary Materials of Ref.~\cite{lee2021multiple}, in~the context of nickel-based pnictides. Here,} we 
only discuss the regimes that may be relevant to our forthcoming
analysis, adopting the simplest possible description of the two
competing phases and their interplay. Thus, for~the sake 
of simplicity, in~the following, we assume that $b$ and $d$ 
are nearly constant (i.e., weakly dependent on the 
external control 
parameters, such as the temperature, the~magnetic field, or~the 
carrier density), and we take $b=d=1$, which amounts 
to rescaling $\Delta$ and $\varphi$. Within~a mean-field
description, one can easily find the various phases of the model
minimizing the free energy in \eqref{eq:GL} with respect to $\Delta$ 
and $\varphi$ and~identifying the conditions for the existence 
and stability of the various phases. The~solutions that make the gradient of the free  energy vanish (extremal points) are 
\begin{eqnarray*}
	(1)&~&\,\Delta=\varphi=0,\qquad \nonumber\\
	(2)&~&\,\begin{cases}
		\Delta^2= -a,  \quad	 a<0,\\
		\varphi=0,
	\end{cases}\qquad
	(3)~~~~~~~~~\,\begin{cases}
		\Delta=0\\
		\varphi^2=-c,  \quad c<0,
	\end{cases}\quad\nonumber\\
	(4)&~&\begin{cases}
		\Delta^2= \frac{a-e c}{e^2-1},\\
		\varphi^2=\frac{c-ea}{e^2-1}.
	\end{cases}
\end{eqnarray*}     

\begin{figure}[h]
	\centering
	\includegraphics[width=0.9\linewidth]{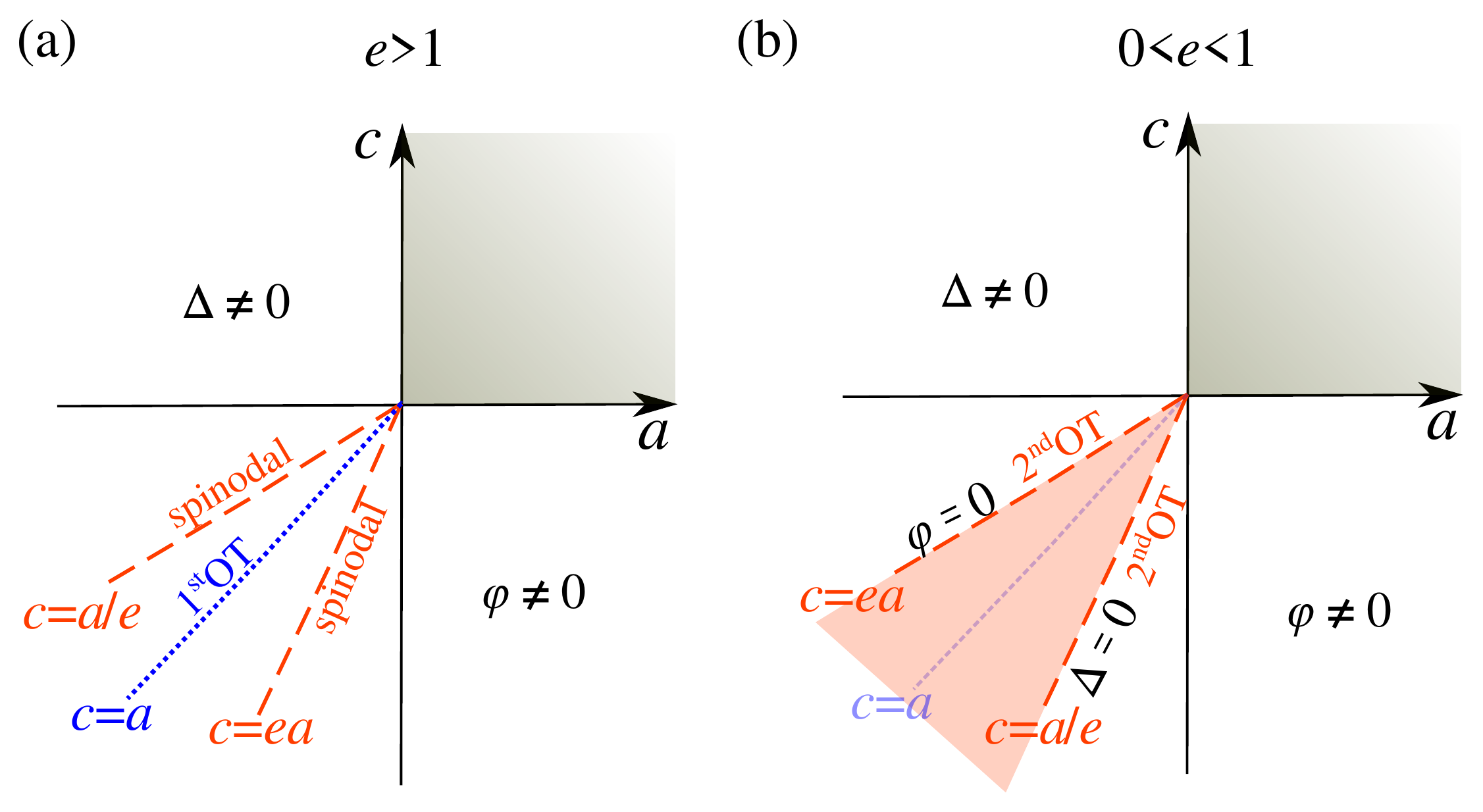}
	\caption{Phase 
		diagram in the $c$ vs. $a$ plane for~the simple model encoded in \eqref{eq:GL}.   (\textbf{a}) $e>1$, the~two
		ordered phases with $\Delta\neq 0$ (stable in the half-plane $a<0$) and with $\varphi\neq 0$ (stable in the half-plane $c<0$)
		are strongly competing. The~half-line $c=a$ represents the first-order transition line between the two phases, but~the two phases can coexist (one as the stable phase, the~other as a metastable phase) in the whole third quadrant of the $c$ vs. $a$ plane, with~the two half-lines $a=0$, $c<0$ and $a<0$, $c=0$, representing the
		spinodal lines for the phases with $\Delta\neq 0$ and 
		$\varphi\neq 0$, respectively;
		(\textbf{b})~$0<e<1$, where instead there is a finite region (highlighted in orange) in which both $\Delta$ and $\varphi$ 
		are simultaneously nonzero. 
		{The gray shaded regions in the first quadrant of both panels represent the disordered phase ($\Delta=\varphi=0$).}}
	\label{fig:GL_phasediag}
\end{figure}

This last solution, with~both order parameters different from 
zero, exists only if $a,c<0$ and either
\begin{equation}
\begin{cases}
0<e<1,\\
a/e<c<ea,
\end{cases}\,\text{or}\quad
\begin{cases}
e>1,\\
ea<c<a/e.
\end{cases}
\end{equation}

{By inserting} 
those solutions in \eqref{eq:GL}, one finds
\begin{equation}
F_1=0,\quad
F_2=-\frac{a^2}{4},\quad
F_3=-\frac{c^2}{4},\quad
F_4=\frac{a^2+c^2 -2 a c e}{4(e^2-1)}.
\end{equation}

{The} nature of various extremum points (minima, maxima, saddle points)
is easily ascertained by inspection of the corresponding Hessian 
matrix.

The solution $\Delta=\varphi=0$ describes the disordered phase. This 
is a minimum of the free energy when both $a$ and $c$ are positive, a~
saddle point if $a$ and $c$ have opposite signs, and~a maximum 
when both $a$ and $c$ are negative. This means that the 
disordered phase is stable in the first quadrant of the $c$
vs. $a$ plane in parameter space. The~phase with 
$\Delta\neq 0$ and $\varphi=0$, describing the
superconducting phase, is a minimum for $a<0$ and $c>ea$;
the phase with $\Delta= 0$ and $\varphi\neq 0$, describing 
the charge-ordered phase, is a minimum for $c<0$ and $c<a/e$.
In the following, we shall call these two phases the pure
phases to~be contrasted with the mixed phase where both
$\Delta$ and $\varphi$ are simultaneously~nonzero.

To discuss the ordered phases of our model in more detail, 
let us consider separately the cases $e>1$ and $0<e<1$.
When $e>1$, we find that $F_{2,3}<F_4$ for all $ea<c<a/e$, 
where the mixed solution with both nonzero order 
parameters exists. 
The half line $c=a$ in the third quadrant of the $c$ vs. $a$ 
plane represents the first-order transition line 
between the two pure phases, but~the two pure phases can 
coexist (one as the stable phase, the~other as a metastable phase) 
in the region of the third quadrant of the $c$ vs. $a$ 
plane encompassed by the two lines $c=ea$ and $c=a/e$, which  
represent the spinodal lines for the phases with $\Delta\neq 0$ 
and $\varphi\neq 0$, respectively {(see Figure~\ref{fig:GL_phasediag}b)}.

Conversely, when $0<e<1$, we find that $F_4<F_{2,3}$ in the
whole region $a/e<c<ea$ of parameter space, 
where the mixed solution with both nonzero order parameters 
exists. In~this latter case, the~mixed phase joins
continuously the pure phase with $\Delta\neq 0$ and $\varphi=0$ 
along the  half-line $c=ea$ in the third quadrant of the $c$ 
vs. $a$ 
plane, and~the pure phase with $\Delta= 0$ and $\varphi\neq 0$ 
along the half-line $c=a/e$ in the third quadrant of the $c$ vs. 
$a$ plane {(see Figure~\ref{fig:GL_phasediag}a)}. Thus, for~positive but sufficiently 
small $e$, the~competition between the two phases is not strong,
and the first-order phase transition between the pure phases 
characterized by either of the two competing orders is 
circumvented by the occurrence of a mixed phase where both 
$\Delta$ and $\varphi$ are~nonzero.

The origin of the phase diagrams in 
Figure~\ref{fig:GL_phasediag} is a {bicritical} point when
$e>1$, and~two second-order critical lines, $\{c=0,\,a>0\}$ 
and $\{a=0,\,c>0\}$ meet with a first-order line, $\{
a=c,\,a,c<0\}$, and~
a {quadricritical} point when $0<e<1$, and~four second-order 
critical lines, $\{c=0,\,a>0\}$ $\{a=0,\,c>0\}$, 
$\{c=ea,\,a,c<0\}$, and~$\{c=a/e,\,a,c<0\}$, meet. In~this latter case, the~mixed phase 
with both $\Delta$ and $\varphi$ different from zero is the
least symmetric of all phases (both symmetries are broken).

{In other words, for~weak competition of the two order parameters, 
	the two symmetries can be simultaneously broken, meaning that the
	two superconducting and charge-density-wave gaps are present at the same
	time and uniformly in the system.
	Conversely, in~the strong competition regime, a (meta)stable phase with 
	both nonzero order parameters cannot exist. The~system exhibits a first-order phase transition between a pure charge-density-wave phase and a pure superconducting phase. In~the region of coexistence of the two minima,  
	a metastable state can nucleate, with~a lifetime that decreases exponentially 
	with the height of the potential barrier and goes to zero in the thermodynamic limit, according to Arrhenius law.  In~absence of any other external source 
	that can stabilize it, the~observation of a metastable state is
	known to be quite difficult. However, disorder can act as a source of 
	nucleation. A~dirty system presenting a strong competition between the two phases
	would hence provide a good playground to observe a phase-separated system that might support, e.g.,~the occurrence of a filamentary superconducting structure.}

We note, on~passing, that the limiting case $e=1$ is peculiar, 
in that it makes the model supersymmetric along the half-line 
$c=a$ in the third quadrant of the $c$ vs. $a$ plane. Along 
this line, the~pure phases with $\Delta\neq 0$ and $\varphi\neq 0$ 
are degenerate to all the phases with the same magnitude of 
$\sqrt{\Delta^2+\varphi^2}$ (this phase might be called 
supersolid).

\section{Competition between superconductivity and charge density waves as a mechanism promoting filamentary superconductivity}
\label{sec:competition}
\begin{figure}[h]
	\centering
	\includegraphics[width=0.95\linewidth]{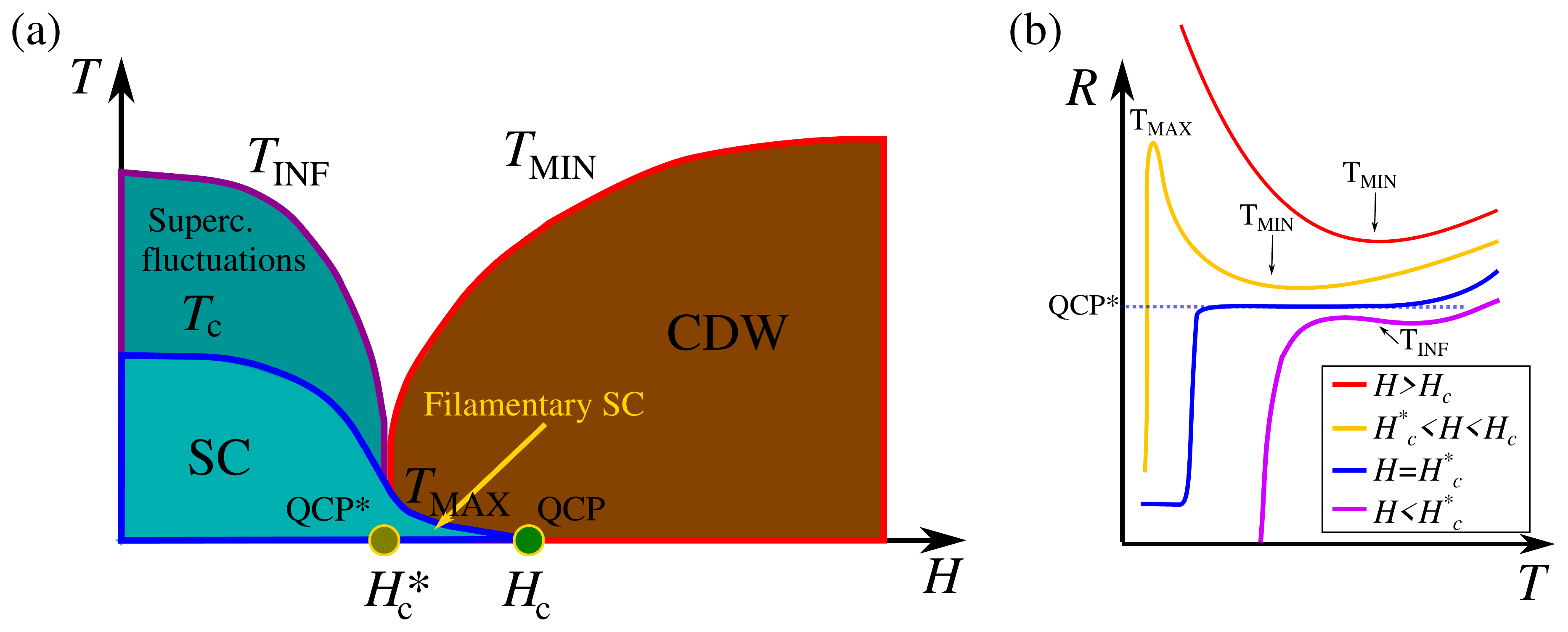}
	\caption{{(\textbf{a}) Phase} 
		diagram of the temperature $T$ vs. magnetic 
		filed $H$ proposed for LSCO based on the characteristic 
		temperature scales identified on the resistance curves 
		measured at various magnetic fields: $T_{\text MIN}(H)$, 
		$T_{\text MAX}(H)$, and~$T_{\text INF}(H)$ are, 
		respectively, the~temperature of the minimum, of~the maximum, 
		and of the inflection point found along the resistance 
		curves, while $T_c$ is the temperature of the
		zero-resistance state. At~zero temperature, two characteristic 
		values of the magnetic field are found, $H_c^*$ and $H_c$, 
		which represent the missed quantum critical point QCP$^*$ (see
		text) and the quantum critical point for the disappearance of
		superconductivity, respectively. Filamentary superconductivity
		is suggested to occur for $H_c^*<H<H_c$. Figure adapted from {Ref}.~\cite{caprara2020doping}, where all the details can be found.
		{(\textbf{b}) Cartoon plot of the resistance measured for various magnetic fields in {Ref}.~\cite{caprara2020doping} and sketched in {Ref}.~\cite{leridon2020protected}.}} 
	\label{fig:phasediag_sketch}
\end{figure}

Let us now discuss the phase diagram proposed for cuprates in 
Ref.~\cite{caprara2020doping}, taking the various 
characteristic temperature scales appearing in the resistance
vs. temperature curves measured in LSCO at various magnetic field{s}
(see Figure~\ref{fig:phasediag_sketch}) as proxies of the various
phases. At~high magnetic fields, the~resistance curves are characterized by a minimum at a temperature 
$T_{\text MIN}(H)$ that decreases with decreasing $H$ 
{(red and yellow curves in panel \textbf{b} show the corresponding resistances)}. This 
temperature scale was proposed to be the proxy of 
charge-density-wave ordering. At~zero or low magnetic field, 
the resistance curves decrease monotonically with decreasing 
the temperature. We take the temperature of the inflection
point $T_{\text INF}(H)$ as the proxy for the onset of superconducting fluctuations, which decreases with increasing $H$ and signals the precursor of superconductivity
{(see also the purple curve in panel {b}
	).}
The two curves $T_{\text MIN}(H)$ and $T_{\text INF}(H)$ meet, with~a vertical
tangent, at~a magnetic field $H^*_c$ that marks 
at zero temperature the quantum critical point  QCP$^*$ for 
the magnetic-field-driven transition between superconductivity
and charge-density wave. {The corresponding resistance, plotted in blue, shows in fact a horizontal tangent signaling the (avoided) quantum critical point, and~then, it drops to zero at some lower temperature $T_{\mathrm c}$.} 
According to the scenario discussed in 
Section~\ref{sec:GL-model}, cuprates appear to be in a regime of
strong competition between the two phases ($e>1$), and the
transition is of first order. The~first-order transition line
is very steep (nearly vertical) in the temperature vs. 
magnetic field plane (see Figure~\ref{fig:phasediag_sketch}).

Surprisingly, however, the~resistance curves 
develop a remarkable non-monotonicity in a window
of magnetic fields $H_c^*<H<H_c$: with decreasing the
temperature, after~reaching the minimum at $T=T_{\text MIN}(H)$,
the resistance curves reach a maximum at $T=T_{\text MAX}(H)$
and then decrease{, as~sketched by the yellow curve of panel {b}}, signaling the appearance of a 
low-temperature superconducting state, surviving {inside the charge-density-wave state} up to
a magnetic field $H_c$ larger than $H^*_c$ (see 
Figure~\ref{fig:phasediag_sketch}).

\begin{figure}[h]
	\centering
	\includegraphics[width=0.55\linewidth]{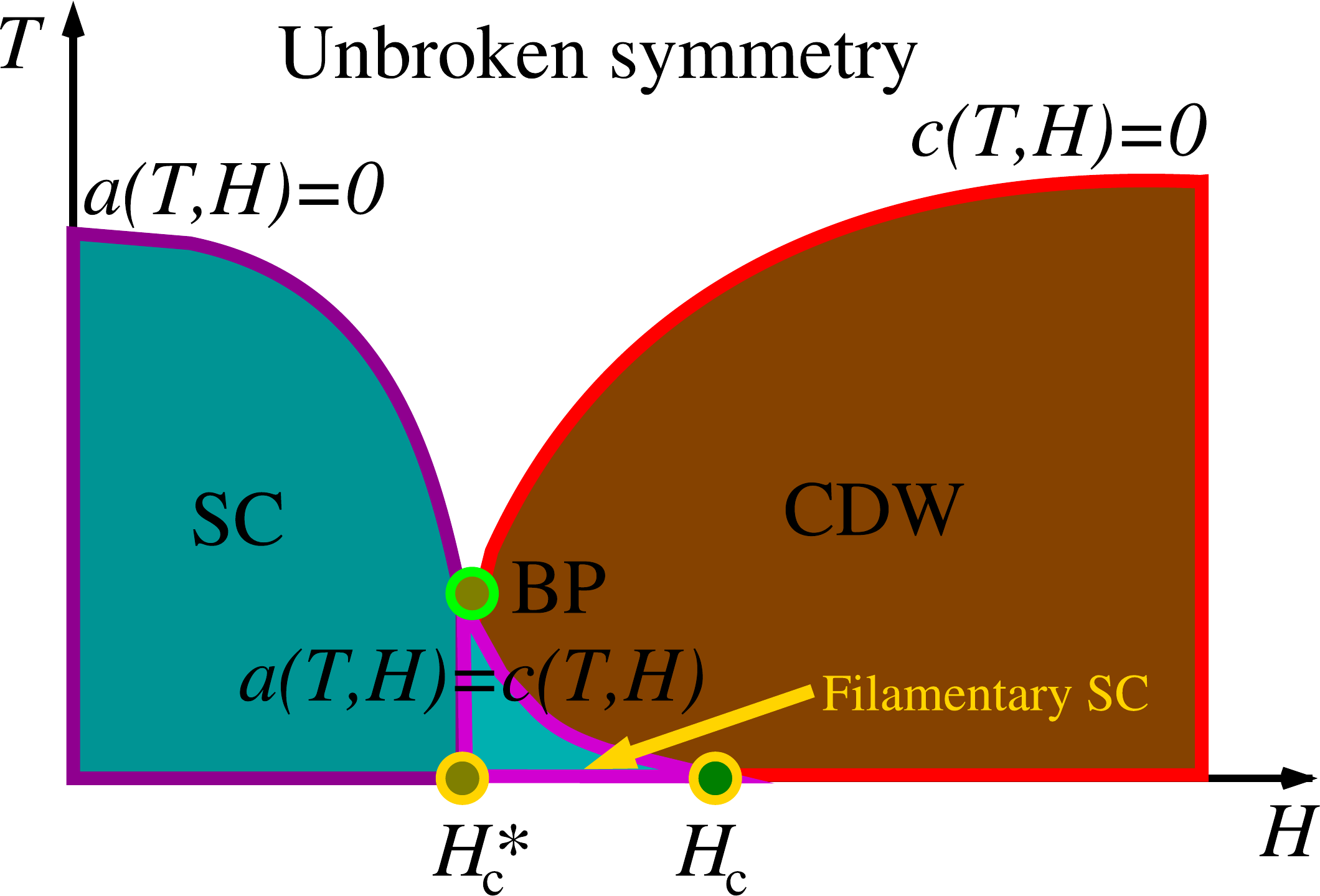}
	\caption{Phase diagram resulting from the comparison of the
		results of the results of Section~\ref{sec:GL-model} and the
		related mean-field phase diagram, 
		Figure~\ref{fig:GL_phasediag}a, with~the phase diagram 
		conjectured in {Ref}.~\cite{caprara2020doping} and displayed 
		in Figure~\ref{fig:phasediag_sketch}. The~label BP marks the bicritical point where the first-order transition line 
		that separates superconductivity and charge-density waves
		meets the two critical lines separating superconductivity 
		and charge-density waves from the unbroken symmetry phase.
		The filamentary superconducting state is highlighted 
		by a lighter shade of cyan.} 
	\label{fig:mf}
\end{figure}

We can compare the above phenomenology with the 
results of Section~\ref{sec:GL-model} in the case $e>1$, which 
crudely captures the competition among superconductivity and charge-density waves. 
The two phases strongly 
compete, which is presumably because the corresponding condensates gain
energy from the same states in momentum space (that can condense 
either in the particle--particle or in the particle--hole channel). 
Once the parameters $a$ and $c$ of
Equation~(\ref{eq:GL}) are assumed to depend on the temperature 
$T$ and on the magnetic field $H$, the~various lines in 
Figure~\ref{fig:GL_phasediag}a translate into lines
in the $T$ vs. $H$ plane. We suggest that the proxy 
for superconductivity,
$T_\text{INF}(H)$ in Figure~\ref{fig:phasediag_sketch} 
corresponds to the mean-field critical line 
described by the implicit equation $a(T,H)=0$,
with $c(T,H)>0$, of~Figure~\ref{fig:GL_phasediag}a; 
see Figure~\ref{fig:mf}. The~proxy for CDW, 
$T_\text{MIN}(H)$ in Figure~\ref{fig:phasediag_sketch} 
corresponds to the mean-field critical line 
described by the implicit equation $c(T,H)=0$,
with $a(T,H)>0$, of~Figure~\ref{fig:GL_phasediag}a; 
see Figure~\ref{fig:mf}. The~two ordered phases meet with the
unbroken symmetry phase, $\Delta=\varphi=0$, at~the
bicritical point $a(T,H)=c(T,H)=0$, which is marked with the symbol
BP in Figure~\ref{fig:mf}. The~bicritical point is the
endpoint of a first-order critical line, 
$a(T,H)=c(T,H)$, with~$a(T,H),\,c(T,H)<0$, which 
in cuprates seems
to be nearly independent of the temperature, terminates
at $T=0$ in correspondence of the magnetic field $H=H_c^*$,
and separates the superconducting and charge-density-wave 
phases. Fluctuations beyond the mean field shrink the ordered
regions of the phase diagram, bringing the superconducting
transition from $T_\text{INF}(H)$ down to $T_c(H)$ and~the 
charge-density-wave transition from $T_\text{MIN}(H)$ down to 
a corresponding transition line $T_\text{CDW}(H)$ (not marked 
in Figure~\ref{fig:phasediag_sketch}).

{The scenario discussed so far does not yet allow for the 
	occurrence of filamentary superconductivity.
	However, in~Refs.\,\cite{caprara2020doping,leridon2020protected},
	it was argued that disorder, which is not taken into account in Equation~(\ref{eq:GL}), might play a crucial role in 
	promoting the onset of filamentary superconductivity, which
	was thereby} conjectured to occur in cuprates for fields
$H_c^*<H<H_c$ as~a {parasitic} phase at sufficiently 
low temperature; see \mbox{Figures~\ref{fig:phasediag_sketch} {and \ref{fig:mf}}}.

Specifically, disorder is expected to mainly couple to
charge-density waves, promoting the fragmentation of the
charge-density-wave state into domains with different
realizations of charge order (in our crude model, regions
with $\varphi>0$ and regions with $\varphi<0$). In~the region 
of the phase diagram where superconductivity exists as a 
metastable phase (the region between the half-line $c=a$ and 
the half-line $c=ea$ in the third quadrant of 
Figure~\ref{fig:GL_phasediag}a) can be
promoted to a locally stable phase within the domain walls
separating the domains with different realizations of 
charge order, possibly giving rise to a 
filamentary superconducting 
state, as~highlighted by a lighter shade of cyan 
in Figure~\ref{fig:mf}.

\begin{figure}[th]
	\centering
	\includegraphics[width=0.9\linewidth]{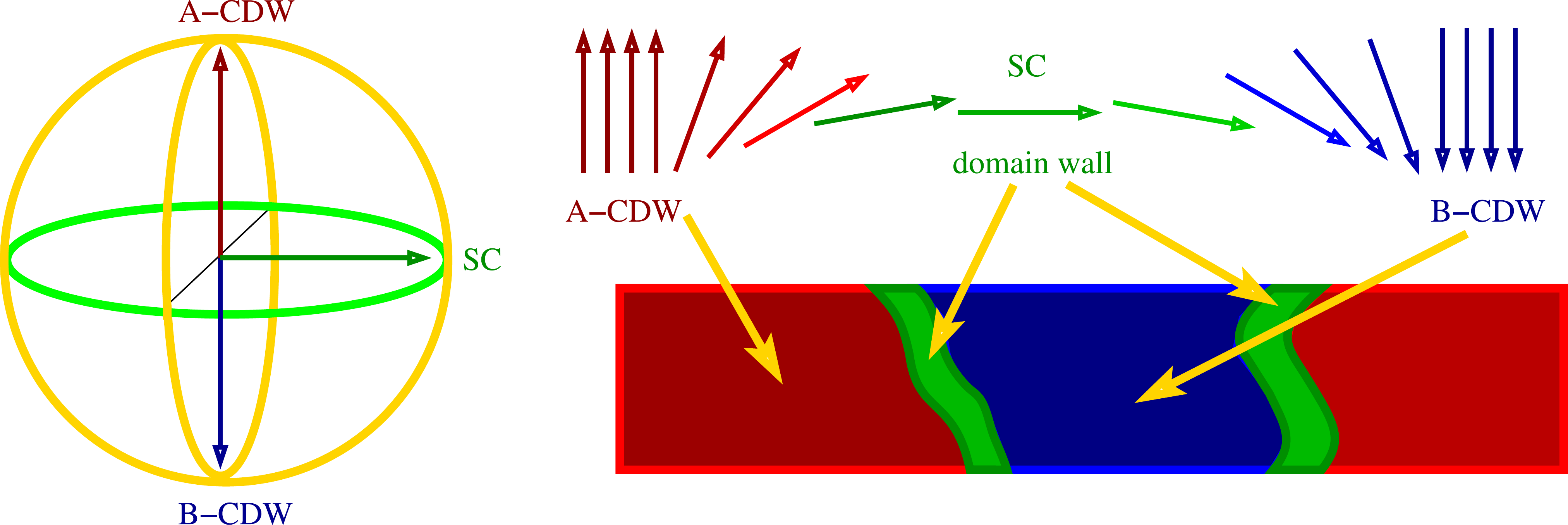}
	\caption{Sketch of the mechanism leading to the occurrence 
		of filamentary superconductivity as a topologically 
		protected  parasitic phase. The~three-dimensional vector $\vec\Psi=(\text{Re}\,\Delta,\text{Im}\,\Delta,\varphi)$,
		is normalized so that its tip can reach any point 
		of a sphere (to the left in the figure). The~two 
		different  realizations of the the charge-density waves 
		(named A-CDW and B-CDW) correspond to the tip 
		of $\vec\Psi$ located at the north or south pole of 
		the sphere, respectively, whereas superconductivity 
		is represented by the tip of $\vec\Psi$ reaching any 
		point along the equator of the sphere. When the vector
		$\vec\Psi$ rotates within a domain wall (to the right 
		in the figure) separating the
		two realizations of charge order (red and blue areas),
		it must necessarily pass through the equator so that
		the domain wall can support a superconducting filament 
		(depicted in green). Figure adapted from {Ref}.~\cite{leridon2020protected}.}
	\label{fig:fsc}
\end{figure}

{To understand why and how the competition between 
	charge-density waves and superconductivity in the presence of disorder
	can trigger the formation of superconducting filaments, we can follow the line
	of reasoning of 
	Ref.~\cite{leridon2020protected}.
	The model introduced there belongs to a class of models
	in which charge-density waves and superconductivity are two manifestations of a (missed) wider symmetry.
	In our Ginzburg--Landau model, as shown in Equation~(\ref{eq:GL}), this wider symmetry is achieved for $e=1$ and $a=c$, 
	as discussed at the end of Section~\ref{sec:GL-model}, and~it is only approximate 
	for $e\neq 1$ and/or $a\neq c$. The~simplest realization of the wider symmetry is $SO(3)$, and it is decomposed into $Z_2\times U(1)$, 
	where $Z_2$ is an Ising-like symmetry describing two different realizations of a commensurate charge-density wave, e.g.,~
	two possible choices for the 
	location of the higher density site in the lattice, and~$U(1)$ is the standard symmetry of the complex order parameter 
	of a superconductor. The~two competing phases were thus encoded in} a three-dimensional vector 
$\vec\Psi=(\text{Re}\,\Delta,\text{Im}\,\Delta,\varphi)$, which is
suitably normalized so that its tip can reach any point of
a sphere (see Figure~\ref{fig:fsc}). The~two different 
realizations of the charge-density waves (named A-CDW and 
B-CDW in Figure~\ref{fig:fsc}) correspond to the tip of 
$\vec\Psi$ located at the north or south pole of the sphere, 
respectively, whereas superconductivity is represented
by the tip of $\vec\Psi$ reaching any point along the equator
of the sphere, as~appropriate to a phase that spontaneously
breaks $U(1)$ symmetry. This description is particularly 
suitable when the statistical mechanical model describing 
the system enjoys approximate $SO(3)$ symmetry in the space
spanned by $\vec\Psi$. When disorder promotes the
fragmentation of the charge-density wave into {neighboring} domains hosting the two different realizations of the charge-density waves, the~vector $\vec\Psi$ must gradually rotate
from the north to the south pole of the sphere. In~doing so,
it needs to cross the equator, hence forming an in-plane domain wall in the form of superconducting filaments. If~the 
filaments form a percolating network endowed with 
superconducting phase coherence (superfluid stiffness), the 
resulting filamentary superconducting
phase can be seen as a topologically protected 
 parasitic phase promoted by disorder~\cite{leridon2020protected}.

{We point out that when discussing the physics of cuprates, 
	we always refer to a single copper--oxygen 
	plane, putting aside the problem of phase coherence between the planes, that eventually results in three-dimensional (bulk) 
	superconductivity. The~occurrence of filamentary superconductivity may somewhat decouple the copper--oxygen planes, but~further work is needed to investigate its connection with the suppression of three-dimensional superconductivity in favor of two-dimensional superconductivity~\cite{li2019tuning}, which is beyond the scope of the present 
	article. 
	It must be borne in mind that a filamentary superconducting state on a single
	copper--oxygen plane is somewhat 
	one-dimensional-like. Still, in~order to have a finite superfluid stiffness, the~superconducting cluster must eventually be connected in a 
	(very loose) two-dimensional structure.} 

\section{Conclusions}
\label{sec:conclusions}

To conclude, we proposed that a competition mechanism among superconductivity 
and charge-density waves can trigger the formation of filamentary 
superconductivity. Experimental evidence in both cuprates and some transition 
metal dichalcogenides indicates a coexistence of the two phases in suitable 
regions of their phase diagram.
In both cases, this competition needs an external control parameter in order
to be triggered. 
In the case of some transition metal dichalcogenides, the~pristine material has a 
commensurate charge-density-wave ground state, which can be destroyed by chemical 
doping, pressure, or~ion-gating techniques, to~drive the appearance of superconductivity.
On the other hand, the~superconducting phase of hole-doped cuprates can be replaced by the charge-density-wave phase through the application of a magnetic field.
Whereas the two materials are indeed very different in many aspects, we proposed to study 
the superconducting vs. charge-density-wave competition by means of a very essential model 
within the Ginzburg--Landau approach, where the two phases are encoded in the parameters $\varphi$ and 
$\Delta$ representing, respectively, the~charge-density-wave and superconducting order parameters.
{We found that for weak competition, i.e.,~$0<e<1$ and 
	$a/e<c<ea$, with~$a,c<0$, $\varphi$ and $\Delta$ are simultaneously and
	uniformly nonzero, i.e., the~two order parameters coexist (Figure~\ref{fig:GL_phasediag}b). In~the case 
	of 
	strong competition, $e>1$,
	while there is no phase with both order parameters different from zero, the~region
	of the phase diagram with $ea<c<a/$ and $a,c<0$ hosts the two phases as local minima
	of the free energy (one stable, the~other metastable) with~the first-order transition 
	line between the two pure phases, $a=c$, with~$a,c<0$, embedded in the coexistence 
	region delimited by the two spinodal lines of 
	Figure~\ref{fig:GL_phasediag}a.}

{In some transition metal dichalcogenides, the~competition with
	an incommensurate charge-density wave may be weak, possibly
	allowing for a phase with both order parameters different from zero.
	The case of strong competition, in~the presence of disorder, may instead be relevant for cuprates
	~\cite{leridon2020protected, caprara2020doping}, in~which a very 
	fragile superconducting 
	state was suggested to be responsible for an avoided quantum critical point and the peculiar behavior of the magnetoresistance curves (see Figure~\ref{fig:phasediag_sketch})}.

As effective as it was, the~model presented in 
Ref.~\cite{leridon2020protected} was missing a {crucial}
part of the discussion, i.e.,~the 
temperature dependence.
One possible way to solve the model at finite temperature is to use Monte Carlo 
simulations to construct a $T$ vs $\alpha$ phase diagram, where $\alpha$ is an external 
(non-thermal) control parameter that allows switching between the two phases. 
Beyond the construction  of a reliable phase diagram at finite temperature, such as~
the one sketched in Figure~\ref{fig:phasediag_sketch}, a~Monte Carlo study can 
also address the temperature behavior of some quantities that better characterize 
the nature of the various phases and transitions.
In particular, a~two-dimensional superconductor undergoes 
a Berezinskii--Kosterlitz--Thouless phase transition 
instead of the phase transition with an order parameter, 
{characterizing the} three-dimensional superconductor, and~the role of phase fluctuations is much more relevant than the usual
Cooper pair fluctuations. In~one-dimensional systems, instead, 
phase slips generated 
by either thermal or quantum fluctuations can break the long-range coherence. The~behavior in temperature of the superfluid stiffness 
would hence 
present different peculiarities depending on whether the superconductor is bulk, 
two-dimensional, homogeneous, inhomogeneous, or~filamentary. 
To assess the presence of a phase coherence, it is thus crucial 
to investigate the temperature behavior of the superfluid stiffness. Such a thorough 
Monte Carlo analysis is currently underway~\cite{venditti2023montecarlo}.

\section*{Author contributions}
Conceptualization, G.V. and S.C.; methodology, G.V. and S.C.; 
formal analysis, G.V. and S.C.; investigation, G.V: and S.C.; writing---original 
draft preparation, G.V. and S.C.; writing---review and editing, G.V. and S.C.; 
supervision, S.C.; funding acquisition, S.C. All authors have read and agreed to the 
published version of the manuscript.

\section*{Funding}
{We acknowledge financial support from the University of Rome Sapienza 
	under the projects Ateneo 2020 (RM120172A8CC7CC7), Ateneo 2021 (RM12117A4A7FD11B), 
	Ateneo 2022 (RM12218162CF9D05), from~the Italian Ministero
	dell’Universit\`a e della Ricerca, under~the Project
	PRIN 2017Z8TS5B, and~from PNRR MUR project~PE0000023-NQSTI.}

\section*{Acknowledgments}
{We acknowledge stimulating discussions with M. Grilli, 
B . Leridon, J. Lorenzana, and I. Maccari.}

%

\begin{thebibliography}{999}
	
	\bibitem{hwang2012emergent}
	Hwang, H.Y.; Iwasa, Y.; Kawasaki, M.; Keimer, B.; Nagaosa, N.; Tokura, Y.
	\newblock {Emergent phenomena at oxide interfaces}.
	\newblock {\em Nat. Mater.} {\bf 2012}, {\em 11},~103--113.
	\newblock {\url{https://doi.org/doi:10.1038/nmat3223}}.
	
	\bibitem{schmidt2015electronic}
	Schmidt, H.; Giustiniano, F.; Eda, G.
	\newblock {Electronic transport properties of transition metal
		dichalcogenide field-effect devices: Surface and interface effects}.
	\newblock {\em Chem. Soc. Rev.} {\bf 2015}, {\em 44},~7715--7736.
	
	\bibitem{rajan2020morphology}
	Rajan, A.; Underwood, K.; Mazzola, F.; King, P.D.C.
	\newblock {Morphology control of epitaxial monolayer transition metal
		dichalcogenides}.
	\newblock {\em Phys. Rev. Mater.} {\bf 2020}, {\em 4},~014003.
	\newblock {\url{https://doi.org/10.1103/PhysRevMaterials.4.014003}}.
	
	\bibitem{yang2023synthesis}
	Yang, R.; Fan, Y.; Mei, L.; Shin, H.S.; Voiry, D.; Lu, Q.; Li, J.; Zeng, Z.
	\newblock {Synthesis of atomically thin sheets by the intercalation-based
		exfoliation of layered materials}.
	\newblock {\em Nat. Synth.} {\bf 2023}, {\em 2},~101--118.
	\newblock {\url{https://doi.org/10.1038/s44160-022-00232-z}}.
	
	\bibitem{geim2013van}
	Geim, A.K.; Grigorieva, I.V.
	\newblock {Van der Waals heterostructures}.
	\newblock {\em Nature} {\bf 2013}, {\em 499},~419--425.
	
	\bibitem{Saito2016}
	Saito, Y.; Nojima, T.; Iwasa, Y.
	\newblock {Highly crystalline 2D superconductors}.
	\newblock {\em Nat. Rev. Mater.} {\bf 2016}, {\em 2},~16094.
	\newblock {\url{https://doi.org/10.1038/natrevmats.2016.94}}.
	
	\bibitem{tsen2016nature}
	Tsen, A.; Hunt, B.; Kim, Y.; Yuan, Z.; Jia, S.; Cava, R.; Hone, J.; Kim, P.;
	Dean, C.; Pasupathy, A.
	\newblock Nature of the quantum metal in a two-dimensional crystalline
	superconductor.
	\newblock {\em Nat. Phys.} {\bf 2016}, {\em 12},~208--212.
	
	\bibitem{kapitulnik2019colloquium}
	Kapitulnik, A.; Kivelson, S.A.; Spivak, B.
	\newblock {Colloquium: Anomalous metals: Failed superconductors}.
	\newblock {\em Rev. Mod. Phys.} {\bf 2019}, {\em 91},~011002.
	\newblock {\url{https://doi.org/10.1103/RevModPhys.91.011002}}.
	
	\bibitem{Dezi2018}
	Dezi, G.; Scopigno, N.; Caprara, S.; Grilli, M.
	\newblock {Negative electronic compressibility and nanoscale inhomogeneity in
		ionic-liquid gated two-dimensional superconductors}.
	\newblock {\em Phys. Rev. B} {\bf 2018}, {\em 98},~214507.
	\newblock {\url{https://doi.org/10.1103/PhysRevB.98.214507}}.
	
	\bibitem{caprara2011effective}
	Caprara, S.; Grilli, M.; Benfatto, L.; Castellani, C.
	\newblock Effective medium theory for superconducting layers: A systematic
	analysis including space correlation effects.
	\newblock {\em Phys. Rev. B} {\bf 2011}, {\em 84},~014514.
	
	\bibitem{tinkham2004introduction}
	Tinkham, M.
	\newblock {\em Introduction to Superconductivity}; Courier Corporation: {Washington, DC, USA,} 
	2004.
	
	\bibitem{biscaras2013multiple}
	Biscaras, J.; Bergeal, N.; Hurand, S.; Feuillet-Palma, C.; Rastogi, A.;
	Budhani, R.; Grilli, M.; Caprara, S.; Lesueur, J.
	\newblock Multiple quantum criticality in a two-dimensional superconductor.
	\newblock {\em Nat. Mater.} {\bf 2013}, {\em 12},~542--548.
	
	\bibitem{bucheli2013metal}
	Bucheli, D.; Caprara, S.; Castellani, C.; Grilli, M.
	\newblock Metal--superconductor transition in low-dimensional superconducting
	clusters embedded in two-dimensional electron systems.
	\newblock {\em New J. Phys.} {\bf 2013}, {\em 15},~023014.
	
	\bibitem{caprara2013multiband}
	Caprara, S.; Biscaras, J.; Bergeal, N.; Bucheli, D.; Hurand, S.;
	Feuillet-Palma, C.; Rastogi, A.; Budhani, R.; Lesueur, J.; Grilli, M.
	\newblock Multiband superconductivity and nanoscale inhomogeneity at oxide
	interfaces.
	\newblock {\em Phys. Rev. B} {\bf 2013}, {\em 88},~020504.
	
	\bibitem{prawiroatmodjo2016evidence}
	Prawiroatmodjo, G.E.; Trier, F.; Christensen, D.V.; Chen, Y.; Pryds, N.;
	Jespersen, T.S.
	\newblock Evidence of weak superconductivity at the room-temperature grown
	LaAlO$_3$/SrTiO$_3$ interface.
	\newblock {\em Phys. Rev. B} {\bf 2016}, {\em 93},~184504.
	
	\bibitem{shen2016observation}
	Shen, S.; Xing, Y.; Wang, P.; Liu, H.; Fu, H.; Zhang, Y.; He, L.; Xie, X.; Lin,
	X.; Nie, J.;  et~al.
	\newblock Observation of quantum griffiths singularity and ferromagnetism at
	the superconducting LaAlO$_3$/SrTiO$_3$(110) interface.
	\newblock {\em Phys. Rev. B} {\bf 2016}, {\em 94},~144517.
	
	\bibitem{Saito2018}
	Saito, Y.; Nojima, T.; Iwasa, Y.
	\newblock {Quantum phase transitions in highly crystalline two-dimensional
		superconductors}.
	\newblock {\em Nat. Commun.} {\bf 2018}, {\em 9},~1--7.
	\newblock {\url{https://doi.org/10.1038/s41467-018-03275-z}}.
	
	\bibitem{caprara2012intrinsic}
	Caprara, S.; Peronaci, F.; Grilli, M.
	\newblock Intrinsic instability of electronic interfaces with strong
	$\mathrm{R}$ashba coupling.
	\newblock {\em Phys. Rev. Lett.} {\bf 2012}, {\em 109},~196401.
	
	\bibitem{scopigno2016phase}
	Scopigno, N.; Bucheli, D.; Caprara, S.; Biscaras, J.; Bergeal, N.; Lesueur, J.;
	Grilli, M.
	\newblock Phase separation from electron confinement at oxide interfaces.
	\newblock {\em Phys. Rev. Lett.} {\bf 2016}, {\em 116},~026804.
	
	\bibitem{caprara2015interplay}
	Caprara, S.; Bergeal, N.; Lesueur, J.; Grilli, M.
	\newblock Interplay between density and superconducting quantum critical
	fluctuations.
	\newblock {\em J. Phys. Condens. Matter} {\bf 2015}, {\em
		27},~425701.
	
	\bibitem{li2021pressure}
	Li, M.; Huang, J.; Guo, W.; Yang, R.; Hu, T.; Yu, A.; Huang, Y.; Zhang, M.;
	Zhang, W.; Zhang, J.M.;  et~al.
	\newblock {Pressure tuning of the iron-based superconductor
		$\mathrm{(Ca_{0.73} La_{0.27}) FeAs_2}$}.
	\newblock {\em Phys. Rev. B} {\bf 2021}, {\em 103},~024502.
	
	\bibitem{xiao2012evidence}
	Xiao, H.; Hu, T.; Dioguardi, A.; Shockley, A.; Crocker, J.; Nisson, D.;
	Viskadourakis, Z.; Tee, X.; Radulov, I.; Almasan, C.;  et~al.
	\newblock {Evidence for filamentary superconductivity nucleated at
		antiphase domain walls in antiferromagnetic $\mathrm{CaFe_2 As_2}$}.
	\newblock {\em Phys. Rev. B} {\bf 2012}, {\em 85},~024530.
	
	\bibitem{xiao2012filamentary}
	Xiao, H.; Hu, T.; He, S.; Shen, B.; Zhang, W.; Xu, B.; He, K.; Han, J.; Singh,
	Y.; Wen, H.;  et~al.
	\newblock {Filamentary superconductivity across the phase diagram of
		$\mathrm{Ba (Fe, Co)_2 As_2}$}.
	\newblock {\em Phys. Rev. B} {\bf 2012}, {\em 86},~064521.
	
	\bibitem{gofryk2014local}
	Gofryk, K.; Pan, M.; Cantoni, C.; Saparov, B.; Mitchell, J.E.; Sefat, A.S.
	\newblock {Local inhomogeneity and filamentary superconductivity in
		$\mathrm{Pr}$-doped $\mathrm{CaFe_2 As_2}$}.
	\newblock {\em Phys. Rev. Lett.} {\bf 2014}, {\em 112},~047005.
	
	\bibitem{machida1989magnetism}
	Machida, K.
	\newblock {Magnetism in La$_2$CuO$_4$ based compounds}.
	\newblock {\em Phys. C Supercond.} {\bf 1989}, {\em 158},~192--196.
	
	\bibitem{kato1990soliton}
	Kato, M.; Machida, K.; Nakanishi, H.; Fujita, M.
	\newblock {Soliton lattice modulation of incommensurate spin density wave
		in two dimensional Hubbard model-a mean field study}.
	\newblock {\em J. Phys. Soc. Jpn.} {\bf 1990}, {\em
		59},~1047--1058.
	
	\bibitem{carlson2004spin}
	Carlson, E.W.; Yao, D.X.; Campbell, D.K.
	\newblock Spin waves in striped phases.
	\newblock {\em Phys. Rev. B} {\bf 2004}, {\em 70},~064505.
	
	\bibitem{carlson2008low}
	Zhao, J.; Yao, D.X.; Li, S.; Hong, T.; Chen, Y.; Chang, S.; Ratcliff, W.; Lynn,
	J.W.; Mook, H.A.; Chen, G.F.;  et~al.
	\newblock Low Energy Spin Waves and Magnetic Interactions in $\mathrm{Sr Fe_2
		As_2 }$.
	\newblock {\em Phys. Rev. Lett.} {\bf 2008}, {\em 101},~167203.
	
	\bibitem{Li2016a}
	Li, L.J.; O'Farrell, E.C.; Loh, K.P.; Eda, G.; {\"{O}}zyilmaz, B.; {Castro
		Neto}, A.H.
	\newblock {Controlling many-body states by the electric-field effect in a
		two-dimensional material}.
	\newblock {\em Nature} {\bf 2016}, {\em 529},~185--189.
	\newblock {\url{https://doi.org/10.1038/nature16175}}.
	
	\bibitem{liao2021coexistence}
	Liao, M.; Wang, H.; Zhu, Y.; Shang, R.; Rafique, M.; Yang, L.; Zhang, H.;
	Zhang, D.; Xue, Q.K.
	\newblock {Coexistence of resistance oscillations and the anomalous metal
		phase in a lithium intercalated TiSe$_2$ superconductor}.
	\newblock {\em Nat. Commun.} {\bf 2021}, {\em 12},~5342.
	\newblock {\url{https://doi.org/10.1038/s41467-021-25671-8}}.
	
	\bibitem{piatti2023superconductivity}
	Piatti, E.; Prando, G.; Meinero, M.; Tresca, C.; Putti, M.; Roddaro, S.;
	Lamura, G.; Shiroka, T.; Carretta, P.; Profeta, G.;  et~al.
	\newblock {Superconductivity induced by gate-driven hydrogen intercalation
		in the charge-density-wave compound 1T $-$ TiSe$_2$}.
	\newblock {\em arXiv} {\bf 2022}, arXiv:2205.12951.
	\newblock {\url{https://doi.org/10.48550/arXiv.2205.12951}}.
	
	\bibitem{spera2019energy}
	Spera, M.; Scarfato, A.; Giannini, E.; Renner, C.
	\newblock Energy-dependent spatial texturing of charge order in 1T $-$ Cu$_x$ TiSe$_2$.
	\newblock {\em Phys. Rev. B} {\bf 2019}, {\em 99},~155133.
	
	\bibitem{burk1991charge}
	Burk, B.; Thomson, R.; Zettl, A.; Clarke, J.
	\newblock Charge-density-wave domains in 1T-TaS 2 observed by satellite
	structure in scanning-tunneling-microscopy images.
	\newblock {\em Phys. Rev. Lett.} {\bf 1991}, {\em 66},~3040.
	
	\bibitem{mcmillan1975time}
	McMillan, W.
	\newblock Time-dependent Laudau theory of charge-density waves in
	transition-metal dichalcogenides.
	\newblock {\em Phys. Rev. B} {\bf 1975}, {\em 12},~1197.
	
	\bibitem{kashiwaya2000tunneling}
	Kashiwaya, S.; Tanaka, Y.
	\newblock Tunnelling effects on surface bound states in unconventional
	superconductors.
	\newblock {\em Rep. Prog. Phys.} {\bf 2000}, {\em 63},~1641.
	\newblock {\url{https://doi.org/10.1088/0034-4885/63/10/202}}.
	
	\bibitem{Joe2014}
	Joe, Y.I.; Chen, X.M.; Ghaemi, P.; Finkelstein, K.D.; {De La Pe{\~{n}}a}, G.A.;
	Gan, Y.; Lee, J.C.; Yuan, S.; Geck, J.; MacDougall, G.J.;  et~al.
	\newblock {Emergence of charge density wave domain walls above the
		superconducting dome in 1T $-$ TiSe$_2$}.
	\newblock {\em Nat. Phys.} {\bf 2014}, {\em 10},~421--425.
	\href{http://xxx.lanl.gov/abs/1309.4051}{{\normalfont [1309.4051]}}.
	\newblock {\url{https://doi.org/10.1038/nphys2935}}.
	
	\bibitem{Kogar2017}
	Kogar, A.; {De La Pena}, G.A.; Lee, S.; Fang, Y.; Sun, S.X.; Lioi, D.B.;
	Karapetrov, G.; Finkelstein, K.D.; Ruff, J.P.; Abbamonte, P.;  et~al.
	\newblock {Observation of a Charge Density Wave Incommensuration Near the
		Superconducting Dome in CuxTiSe$_2$}.
	\newblock {\em Phys. Rev. Lett.} {\bf 2017}, {\em 118},~1--5.
	\newblock {\url{https://doi.org/10.1103/PhysRevLett.118.027002}}.
	
	\bibitem{liu2013superconductivity}
	Liu, Y.; Ang, R.; Lu, W.; Song, W.; Li, L.; Sun, Y.
	\newblock Superconductivity induced by Se-doping in layered charge-density-wave
	system 1T $-$ TaS$_{2-x}$Se$x$.
	\newblock {\em Appl. Phys. Lett.} {\bf 2013}, {\em 102},~192602.
	
	\bibitem{Sipos2008b}
	Sipos, B.; Kusmartseva, A.F.; Akrap, A.; Berger, H.; Forr{\'{o}}, L.; Tutis, E.
	\newblock {From Mott state to superconductivity in 1T-TaS$_2$.}
	\newblock {\em Nat. Mater.} {\bf 2008}, {\em 7},~960--965.
	\newblock {\url{https://doi.org/10.1038/nmat2318}}.
	
	\bibitem{mutka1983superconductivity}
	Mutka, H.
	\newblock Superconductivity in irradiated charge-density-wave compounds 2\emph{H} $-$ NbSe$_2$, 2\emph{H} $-$ TaS$_2$, and 2\emph{H} $-$ TaSe$_2$.
	\newblock {\em Phys. Rev. B} {\bf 1983}, {\em 28},~2855.
	
	\bibitem{qian2007emergence}
	Qian, D.; Hsieh, D.; Wray, L.; Morosan, E.; Wang, N.; Xia, Y.; Cava, R.; Hasan,
	M.
	\newblock Emergence of Fermi pockets in a new excitonic charge-density-wave
	melted superconductor.
	\newblock {\em Phys. Rev. Lett.} {\bf 2007}, {\em 98},~117007.
	
	\bibitem{di1976electronic}
	Di~Salvo, F.J.; Moncton, D.; Waszczak, J.
	\newblock Electronic properties and superlattice formation in the semimetal
	TiSe$_2$.
	\newblock {\em Phys. Rev. B} {\bf 1976}, {\em 14},~4321.
	
	\bibitem{kidd2002electron}
	Kidd, T.; Miller, T.; Chou, M.; Chiang, T.C.
	\newblock Electron-hole coupling and the charge density wave transition in TiSe$_2$.
	\newblock {\em Phys. Rev. Lett.} {\bf 2002}, {\em 88},~226402.
	
	\bibitem{cercellier2007evidence}
	Cercellier, H.; Monney, C.; Clerc, F.; Battaglia, C.; Despont, L.; Garnier, M.;
	Beck, H.; Aebi, P.; Patthey, L.; Berger, H.;  et~al.
	\newblock Evidence for an Excitonic Insulator Phase in 1T $-$ TiSe$_2$.
	\newblock {\em Phys. Rev. Lett.} {\bf 2007}, {\em 99},~146403.
	
	\bibitem{hellmann2012time}
	Hellmann, S.; Rohwer, T.; Kall{\"a}ne, M.; Hanff, K.; Sohrt, C.; Stange, A.;
	Carr, A.; Murnane, M.; Kapteyn, H.; Kipp, L.;  et~al.
	\newblock Time-domain classification of charge-density-wave insulators.
	\newblock {\em Nat. Commun.} {\bf 2012}, {\em 3},~1069.
	
	\bibitem{rohwer2011collapse}
	Rohwer, T.; Hellmann, S.; Wiesenmayer, M.; Sohrt, C.; Stange, A.; Slomski, B.;
	Carr, A.; Liu, Y.; Avila, L.M.; Kall{\"a}ne, M.;  et~al.
	\newblock Collapse of long-range charge order tracked by time-resolved
	photoemission at high momenta.
	\newblock {\em Nature} {\bf 2011}, {\em 471},~490--493.
	
	\bibitem{Rossnagel2011}
	Rossnagel, K.
	\newblock {On the origin of charge-density waves in select layered
		transition-metal dichalcogenides}.
	\newblock {\em J. Phys. Condens. Matter} {\bf 2011}, {\em 23}, {213001}.
	 \href{http://xxx.lanl.gov/abs/0402594v3}{{\normalfont
	 [arXiv:arXiv:cond-mat/0402594v3]}}.
	\newblock {\url{https://doi.org/10.1088/0953-8984/23/21/213001}}.
	
	\bibitem{weber2011electron}
	Weber, F.; Rosenkranz, S.; Castellan, J.P.; Osborn, R.; Karapetrov, G.; Hott,
	R.; Heid, R.; Bohnen, K.P.; Alatas, A.
	\newblock Electron-phonon coupling and the soft phonon mode in TiSe$ _2$.
	\newblock {\em Phys. Rev. Lett.} {\bf 2011}, {\em 107},~266401.
	
	\bibitem{Porer2014}
	Porer, M.; Leierseder, U.; M{\'{e}}nard, J.M.; Dachraoui, H.; Mouchliadis, L.;
	Perakis, I.E.; Heinzmann, U.; Demsar, J.; Rossnagel, K.; Huber, R.
	\newblock {Non-thermal separation of electronic and structural orders in a
		persisting charge density wave}.
	\newblock {\em Nat. Mater.} {\bf 2014}, {\em 13},~857--861.
	\newblock {\url{https://doi.org/10.1038/nmat4042}}.
	
	\bibitem{van2010alternative}
	van Wezel, J.; Nahai-Williamson, P.; Saxena, S.S.
	\newblock An alternative interpretation of recent ARPES measurements on TiSe$_2$.
	\newblock {\em Europhys. Lett.} {\bf 2010}, {\em 89},~47004.
	
	\bibitem{van2010exciton}
	van Wezel, J.; Nahai-Williamson, P.; Saxena, S.S.
	\newblock Exciton-phonon-driven charge density wave in TiSe$ _2$.
	\newblock {\em Phys. Rev. B} {\bf 2010}, {\em 81},~165109.
	
	\bibitem{novello2017stripe}
	Novello, A.M.; Spera, M.; Scarfato, A.; Ubaldini, A.; Giannini, E.; Bowler, D.;
	Renner, C.
	\newblock Stripe and Short Range Order in the Charge Density Wave of 1T $-$ Cu$x$TiSe$ _2$.
	\newblock {\em Phys. Rev. Lett.} {\bf 2017}, {\em 118},~017002.
	
	\bibitem{yan2017influence}
	Yan, S.; Iaia, D.; Morosan, E.; Fradkin, E.; Abbamonte, P.; Madhavan, V.
	\newblock Influence of Domain Walls in the Incommensurate Charge Density Wave
	State of Cu Intercalated 1T $-$ TiSe$ _2$.
	\newblock {\em Phys. Rev. Lett.} {\bf 2017}, {\em 118},~106405.
	
	\bibitem{novello2015scanning}
	Novello, A.M.; Hildebrand, B.; Scarfato, A.; Didiot, C.; Monney, G.; Ubaldini,
	A.; Berger, H.; Bowler, D.; Aebi, P.; Renner, C.
	\newblock Scanning tunneling microscopy of the charge density wave in 1T $-$ TiSe$ _2$ in the presence of single atom defects.
	\newblock {\em Phys. Rev. B} {\bf 2015}, {\em 92},~081101.
	
	\bibitem{hildebrand2014doping}
	Hildebrand, B.; Didiot, C.; Novello, A.M.; Monney, G.; Scarfato, A.; Ubaldini,
	A.; Berger, H.; Bowler, D.; Renner, C.; Aebi, P.
	\newblock Doping Nature of Native Defects in 1T $-$ TiSe$ _2$.
	\newblock {\em Phys. Rev. Lett.} {\bf 2014}, {\em 112},~197001.
	
	\bibitem{dai2014microscopic}
	Dai, J.; Calleja, E.; Alldredge, J.; Zhu, X.; Li, L.; Lu, W.; Sun, Y.; Wolf,
	T.; Berger, H.; McElroy, K.
	\newblock Microscopic evidence for strong periodic lattice distortion in
	two-dimensional charge-density wave systems.
	\newblock {\em Phys. Rev. B} {\bf 2014}, {\em 89},~165140.
	
	\bibitem{costanzo2016gate}
	Costanzo, D.; Jo, S.; Berger, H.; Morpurgo, A.F.
	\newblock Gate-induced superconductivity in atomically thin MoS2 crystals.
	\newblock {\em Nat. Nanotechnol.} {\bf 2016}, {\em 11},~339--344.
	
	\bibitem{petach2017disorder}
	Petach, T.A.; Reich, K.V.; Zhang, X.; Watanabe, K.; Taniguchi, T.; Shklovskii,
	B.I.; Goldhaber-Gordon, D.
	\newblock Disorder from the bulk ionic liquid in electric double layer
	transistors.
	\newblock {\em ACS Nano} {\bf 2017}, {\em 11},~8395--8400.
	
	\bibitem{Timusk1999b}
	Timusk, T.
	\newblock {Infrared properties of exotic superconductors}.
	\newblock {\em Phys. C Supercond. Its Appl.} {\bf 1999}, {\em 317--318},~18--29.
	\newblock {\url{https://doi.org/10.1016/S0921-4534(99)00042-8}}.
	
	\bibitem{Fauque2006}
	Fauqu{\'{e}}, B.; Sidis, Y.; Hinkov, V.; Pailh{\`{e}}s, S.; Lin, C.T.; Chaud,
	X.; Bourges, P.
	\newblock {Magnetic order in the pseudogap phase of high-Tc superconductors.}
	\newblock {\em Phys. Rev. Lett.} {\bf 2006}, {\em 96},~197001.
	\newblock {\url{https://doi.org/10.1103/PhysRevLett.96.197001}}.
	
	\bibitem{mangin-thro2017ab}
	Mangin-Thro, L.; Li, Y.; Sidis, Y.; Bourges, P.
	\newblock $a\ensuremath{-}b$ Anisotropy of the Intra-Unit-Cell Magnetic Order
	in ${\mathrm{YBa}}_{2}{\mathrm{Cu}}_{3}{\mathrm{O}}_{6.6}$.
	\newblock {\em Phys. Rev. Lett.} {\bf 2017}, {\em 118},~097003.
	\newblock {\url{https://doi.org/10.1103/PhysRevLett.118.097003}}.
	
	\bibitem{mook2008observation}
	Mook, H.A.; Sidis, Y.; Fauqu\'e, B.; Bal\'edent, V.; Bourges, P.
	\newblock Observation of magnetic order in a superconducting
	${\text{YBa}}_{2}{\text{Cu}}_{3}{\text{O}}_{6.6}$ single crystal using
	polarized neutron scattering.
	\newblock {\em Phys. Rev. B} {\bf 2008}, {\em 78},~020506.
	\newblock {\url{https://doi.org/10.1103/PhysRevB.78.020506}}.
	
	\bibitem{baledent2011evidence}
	Bal\'edent, V.; Haug, D.; Sidis, Y.; Hinkov, V.; Lin, C.T.; Bourges, P.
	\newblock Evidence for competing magnetic instabilities in underdoped
	YBa${}_{2}$Cu${}_{3}$O${}_{6+x}$.
	\newblock {\em Phys. Rev. B} {\bf 2011}, {\em 83},~104504.
	\newblock {\url{https://doi.org/10.1103/PhysRevB.83.104504}}.
	
	\bibitem{leridon2007josephson}
	Leridon, B.; Ng, T.K.; Varma, C.M.
	\newblock Josephson Effect for Superconductors Lacking Time-Reversal and
	Inversion Symmetries.
	\newblock {\em Phys. Rev. Lett.} {\bf 2007}, {\em 99},~027002.
	\newblock {\url{https://doi.org/10.1103/PhysRevLett.99.027002}}.
	
	\bibitem{baledent2010twodimensional}
	Bal\'edent, V.; Fauqu\'e, B.; Sidis, Y.; Christensen, N.B.; Pailh\`es, S.;
	Conder, K.; Pomjakushina, E.; Mesot, J.; Bourges, P.
	\newblock Two-Dimensional Orbital-Like Magnetic Order in the High-Temperature
	${\mathrm{La}}_{2\ensuremath{-}x}{\mathrm{Sr}}_{x}{\mathrm{CuO}}_{4}$
	Superconductor.
	\newblock {\em Phys. Rev. Lett.} {\bf 2010}, {\em 105},~027004.
	\newblock {\url{https://doi.org/10.1103/PhysRevLett.105.027004}}.
	
	\bibitem{dealmeida2012evidence}
	De~Almeida-Didry, S.; Sidis, Y.; Bal\'edent, V.; Giovannelli, F.; Monot-Laffez,
	I.; Bourges, P.
	\newblock Evidence for intra-unit-cell magnetic order in
	Bi${}_{2}$Sr${}_{2}$CaCu${}_{2}$O${}_{8+\ensuremath{\delta}}$.
	\newblock {\em Phys. Rev. B} {\bf 2012}, {\em 86},~020504.
	\newblock {\url{https://doi.org/10.1103/PhysRevB.86.020504}}.
	
	\bibitem{mangin-thro2014characterization}
	Mangin-Thro, L.; Sidis, Y.; Bourges, P.; De~Almeida-Didry, S.; Giovannelli, F.;
	Laffez-Monot, I.
	\newblock Characterization of the intra-unit-cell magnetic order in
	Bi${}_{2}$Sr${}_{2}$CaCu${}_{2}$O${}_{8+\ensuremath{\delta}}$.
	\newblock {\em Phys. Rev. B} {\bf 2014}, {\em 89},~094523.
	\newblock {\url{https://doi.org/10.1103/PhysRevB.89.094523}}.
	
	\bibitem{li2011magnetic}
	Li, Y.; Bal\'edent, V.; Bari\ifmmode \check{s}\else
	\v{s}\fi{}i\ifmmode~\acute{c}\else \'{c}\fi{}, N.; Cho, Y.C.; Sidis, Y.; Yu,
	G.; Zhao, X.; Bourges, P.; Greven, M.
	\newblock Magnetic order in the pseudogap phase of
	HgBa${}_{2}$CuO${}_{4+\ensuremath{\delta}}$ studied by spin-polarized neutron
	diffraction.
	\newblock {\em Phys. Rev. B} {\bf 2011}, {\em 84},~224508.
	\newblock {\url{https://doi.org/10.1103/PhysRevB.84.224508}}.
	
	\bibitem{tang2018orientation}
	Tang, Y.; Mangin-Thro, L.; Wildes, A.; Chan, M.K.; Dorow, C.J.; Jeong, J.;
	Sidis, Y.; Greven, M.; Bourges, P.
	\newblock Orientation of the intra-unit-cell magnetic moment in the
	high-${T}_{\mathrm{c}}$ superconductor
	$\mathrm{HgB}{\mathrm{a}}_{2}\mathrm{Cu}{\mathrm{O}}_{4+\ensuremath{\delta}}$.
	\newblock {\em Phys. Rev. B} {\bf 2018}, {\em 98},~214418.
	\newblock {\url{https://doi.org/10.1103/PhysRevB.98.214418}}.
	
	\bibitem{doiron2007quantum}
	Doiron-Leyraud, N.; Proust, C.; LeBoeuf, D.; Levallois, J.; Bonnemaison, J.B.;
	Liang, R.; Bonn, D.; Hardy, W.; Taillefer, L.
	\newblock Quantum oscillations and the Fermi surface in an underdoped high-\emph{T}$_\text{c}$
	superconductor.
	\newblock {\em Nature} {\bf 2007}, {\em 447},~565--568.
	\newblock {\url{https://doi.org/10.1038/nature05872}}.
	
	\bibitem{doiron2015evidence}
	Doiron-Leyraud, N.; Badoux, S.; Ren{\'e}~de Cotret, S.; Lepault, S.; LeBoeuf,
	D.; Lalibert{\'e}, F.; Hassinger, E.; Ramshaw, B.; Bonn, D.; Hardy, W.;
	et~al.
	\newblock Evidence for a small hole pocket in the Fermi surface of underdoped
	$\mathrm{YBa_2Cu_3O}_y$.
	\newblock {\em Nat. Commun.} {\bf 2015}, {\em 6},~6034.
	\newblock {\url{https://doi.org/10.1038/ncomms7034}}.
	
	\bibitem{sebastian2014normal}
	Sebastian, S.E.; Harrison, N.; Balakirev, F.; Altarawneh, M.; Goddard, P.;
	Liang, R.; Bonn, D.; Hardy, W.; Lonzarich, G.
	\newblock Normal-state nodal electronic structure in underdoped high-\emph{T}$_\text{c}$ copper
	oxides.
	\newblock {\em Nature} {\bf 2014}, {\em 511},~61--64.
	\newblock {\url{https://doi.org/10.1038/nature13326}}.
	
	\bibitem{ramshaw2015quasiparticle}
	Ramshaw, B.; Sebastian, S.; McDonald, R.; Day, J.; Tan, B.; Zhu, Z.; Betts, J.;
	Liang, R.; Bonn, D.; Hardy, W.;  et~al.
	\newblock Quasiparticle mass enhancement approaching optimal doping in a high-\emph{T}$_\text{c}$ superconductor.
	\newblock {\em Science} {\bf 2015}, {\em 348},~317--320.
	\newblock {\url{https://doi.org/10.1126/science.aaa4990}}.
	
	\bibitem{badoux2016change}
	Badoux, S.; Tabis, W.; Lalibert{\'e}, F.; Grissonnanche, G.; Vignolle, B.;
	Vignolles, D.; B{\'e}ard, J.; Bonn, D.; Hardy, W.; Liang, R.;  et~al.
	\newblock Change of carrier density at the pseudogap critical point of a
	cuprate superconductor.
	\newblock {\em Nature} {\bf 2016}, {\em 531},~210--214.
	\newblock {\url{https://doi.org/10.1038/nature16983}}.
	
	\bibitem{lorenzana1993optical}
	Lorenzana, J.; Yu, L.
	\newblock Optical conductivity of
	${\mathrm{La}}_{2\mathrm{\ensuremath{-}}\mathit{x}}$${\mathrm{Sr}}_{\mathit{x}}$${\mathrm{CuO}}_{4}$
	and soft electronic modes.
	\newblock {\em Phys. Rev. Lett.} {\bf 1993}, {\em 70},~861--864.
	\newblock {\url{https://doi.org/10.1103/PhysRevLett.70.861}}.
	
	\bibitem{wu2012charge}
	Wu, H.H.; Buchholz, M.; Trabant, C.; Chang, C.; Komarek, A.; Heigl, F.;
	Zimmermann, M.; Cwik, M.; Nakamura, F.; Braden, M.;  et~al.
	\newblock Charge stripe order near the surface of 12-percent doped
	$\mathrm{La_{2- x} Sr_x CuO_4}$.
	\newblock {\em Nat. Commun.} {\bf 2012}, {\em 3},~1--5.
	
	\bibitem{Gerber2015}
	Gerber, S.; Jang, H.; Nojiri, H.; Matsuzawa, S.; Yasumura, H.; Bonn, D.A.;
	Liang, R.; Hardy, W.N.; Islam, Z.; Mehta, A.;  et~al.
	\newblock {Three-dimensional charge density wave order in YBa$_2$Cu$_3$O$_{6.67}$ at
		high magnetic fields}.
	\newblock {\em Science} {\bf 2015}, {\em 350},~949--952.
	\newblock {\url{https://doi.org/10.1126/science.aac6257}}.
	
	\bibitem{arpaia2019dynamical}
	Arpaia, R.; Caprara, S.; Fumagalli, R.; De~Vecchi, G.; Peng, Y.; Andersson, E.;
	Betto, D.; De~Luca, G.; Brookes, N.; Lombardi, F.;  et~al.
	\newblock Dynamical charge density fluctuations pervading the phase diagram of
	a $\mathrm{Cu}$-based high-${T_\text{c}}$ superconductor.
	\newblock {\em Science} {\bf 2019}, {\em 365},~906--910.
	
	\bibitem{keimer2015quantum}
	Keimer, B.; Kivelson, S.A.; Norman, M.R.; Uchida, S.; Zaanen, J.
	\newblock From quantum matter to high-temperature superconductivity in copper
	oxides.
	\newblock {\em Nature} {\bf 2015}, {\em 518},~179--186.
	
	\bibitem{comin2016resonant}
	Comin, R.; Damascelli, A.
	\newblock Resonant X-ray Scattering Studies of Charge Order in Cuprates.
	\newblock {\em Annu. Rev. Condens. Matter Phys.} {\bf 2016}, {\em
		7},~369--405.
	\newblock {\url{https://doi.org/10.1146/annurev-conmatphys-031115-011401}}.
	
	\bibitem{Miao2017}
	Miao, H.; Lorenzana, J.; Seibold, G.; Peng, Y.Y.; Amorese, A.; Yakhou-Harris,
	F.; Kummer, K.; Brookes, N.B.; Konik, R.M.; Thampy, V.;  et~al.
	\newblock {High-temperature charge density wave correlations in La$_{1.875}$Ba$_{0.125}$CuO$_4$ without spin–charge locking}.
	\newblock {\em Proc. Natl. Acad. Sci. USA} {\bf 2017}, {\em
		114},~12430--12435.
	\newblock {\url{https://doi.org/10.1073/pnas.1708549114}}.
	
	\bibitem{peng2017}
	Peng, Y.Y.; Dellea, G.; Minola, M.; Conni, M.; Amorese, A.; {Di Castro}, D.;
	{De Luca}, G.M.; Kummer, K.; Salluzzo, M.; Sun, X.;  et~al.
	\newblock {Influence of apical oxygen on the extent of in-plane exchange
		interaction in cuprate superconductors}.
	\newblock {\em Nat. Phys.} {\bf 2017}, {\em 13},~1201--1206.
	\newblock {\url{https://doi.org/10.1038/nphys4248}}.
	
	\bibitem{miao2019formation}
	Miao, H.; Fumagalli, R.; Rossi, M.; Lorenzana, J.; Seibold, G.; Yakhou-Harris,
	F.; Kummer, K.; Brookes, N.B.; Gu, G.D.; Braicovich, L.;  et~al.
	\newblock Formation of Incommensurate Charge Density Waves in Cuprates.
	\newblock {\em Phys. Rev. X} {\bf 2019}, {\em 9},~031042.
	\newblock {\url{https://doi.org/10.1103/PhysRevX.9.031042}}.
	
	\bibitem{wu2015incipient}
	Wu, T.; Mayaffre, H.; Kr{\"a}mer, S.; Horvati{\'c}, M.; Berthier, C.; Hardy,
	W.; Liang, R.; Bonn, D.; Julien, M.H.
	\newblock Incipient charge order observed by NMR in the normal state of
	$\mathrm{YBa_2Cu_3O}_y$.
	\newblock {\em Nat. Commun.} {\bf 2015}, {\em 6},~6438.
	\newblock {\url{https://doi.org/10.1038/ncomms7438}}.
	
	\bibitem{sachdev2016novel}
	Sachdev, S.; Chowdhury, D.
	\newblock {The novel metallic states of the cuprates: Topological Fermi liquids
		and strange metals}.
	\newblock {\em Prog. Theor. Exp. Phys.} {\bf 2016},
	{\em 2016}, 12C102.
	 \href{http://xxx.lanl.gov/abs/https://academic.oup.com/ptep/article-pdf/2016/12/12C102/9620269/ptw110.pdf}{{\normalfont
	[https://academic.oup.com/ptep/article-pdf/2016/12/12C102/9620269/ptw110.pdf]}}.
	\newblock {\url{https://doi.org/10.1093/ptep/ptw110}}.
	
	\bibitem{Varma1999}
	Varma, C.M.
	\newblock {Pseudogap Phase and the Quantum-Critical Point in Copper-Oxide
		Metals}.
	\newblock {\em Phys. Rev. Lett.} {\bf 1999}, {\em 83},~3538--3541.
	\newblock {\url{https://doi.org/10.1103/PhysRevLett.83.3538}}.
	
	\bibitem{varma2006theory}
	Varma, C.M.
	\newblock Theory of the pseudogap state of the cuprates.
	\newblock {\em Phys. Rev. B} {\bf 2006}, {\em 73},~155113.
	\newblock {\url{https://doi.org/10.1103/PhysRevB.73.155113}}.
	
	\bibitem{Zaanen1989a}
	Zaanen, J.; Gunnarsson, O.
	\newblock {Charged magnetic domain lines and the magnetism of high-\emph{T}$_\text{c}$
		oxides}.
	\newblock {\em Phys. Rev. B} {\bf 1989}, {\em 40},~7391--7394.
	\newblock {\url{https://doi.org/10.1103/PhysRevB.40.7391}}.
	
	\bibitem{Castellani1995}
	Castellani, C.; Castro, C.D.; Grilli, M.
	\newblock {Singular quasiparticle scattering in the proximity of charge
		instabilities}.
	\newblock {\em Phys. Rev. Lett.} {\bf 1995}, {\em 75},~4650.
	\newblock {\url{https://doi.org/10.1103/physrevlett.75.4650}}.
	
	\bibitem{lorenzana2002}
	Lorenzana, J.; Castellani, C.; {Di Castro}, C.
	\newblock {Curie temperature and frustrated phase separation in manganites}.
	\newblock {\em Phys. B Condens. Matter} {\bf 2002}, {\em 320},~56--59.
	\newblock {\url{https://doi.org/10.1016/S0921-4526(02)00619-1}}.
	
	\bibitem{castellani1996non}
	Castellani, C.; Di~Castro, C.; Grilli, M.
	\newblock Non-Fermi-liquid behavior and d-wave superconductivity near the
	charge-density-wave quantum critical point.
	\newblock {\em Z. F{\"u}r Phys. B Condens. Matter} {\bf 1996}, {\em
		103},~137--144.
	\newblock {\url{https://doi.org/10.1007/s002570050347}}.
	
	\bibitem{kivelson2003howto}
	Kivelson, S.A.; Bindloss, I.P.; Fradkin, E.; Oganesyan, V.; Tranquada, J.M.;
	Kapitulnik, A.; Howald, C.
	\newblock How to detect fluctuating stripes in the high-temperature
	superconductors.
	\newblock {\em Rev. Mod. Phys.} {\bf 2003}, {\em 75},~1201--1241.
	\newblock {\url{https://doi.org/10.1103/RevModPhys.75.1201}}.
	
	\bibitem{caprara2017dynamical}
	Caprara, S.; Di~Castro, C.; Seibold, G.; Grilli, M.
	\newblock Dynamical charge density waves rule the phase diagram of cuprates.
	\newblock {\em Phys. Rev. B} {\bf 2017}, {\em 95},~224511.
	\newblock {\url{https://doi.org/10.1103/PhysRevB.95.224511}}.
	
	\bibitem{grissonnanche2014direct}
	Grissonnanche, G.; Cyr-Choini{\`e}re, O.; Lalibert{\'e}, F.; Ren{\'e}~de
	Cotret, S.; Juneau-Fecteau, A.; Dufour-Beaus{\'e}jour, S.; Delage, M.E.;
	LeBoeuf, D.; Chang, J.; Ramshaw, B.;  et~al.
	\newblock Direct measurement of the upper critical field in cuprate
	superconductors.
	\newblock {\em Nat. Commun.} {\bf 2014}, {\em 5},~3280.
	\newblock {\url{https://doi.org/10.1038/ncomms4280}}.
	
	\bibitem{caprara2020doping}
	Caprara, S.; Grilli, M.; Lorenzana, J.; Leridon, B.
	\newblock Doping-dependent competition between superconductivity and
	polycrystalline charge density waves.
	\newblock {\em SciPost Phys.} {\bf 2020}, {\em 8},~003.
	
	\bibitem{abanov2003quantum}
	Abanov, A.; Chubukov, A.V.; Schmalian, J.
	\newblock Quantum-critical theory of the spin-fermion model and its application
	to cuprates: Normal state analysis.
	\newblock {\em Adv. Phys.} {\bf 2003}, {\em 52},~119--218.
	\newblock {\url{https://doi.org/10.1080/0001873021000057123}}.
	
	\bibitem{leridon2007paraconductivity}
	Leridon, B.; Vanacken, J.; Wambecq, T.; Moshchalkov, V.V.
	\newblock Paraconductivity of underdoped
	${\mathrm{La}}_{2\ensuremath{-}x}{\mathrm{Sr}}_{x}\mathrm{Cu}{\mathrm{O}}_{4}$
	thin-film superconductors using high magnetic fields.
	\newblock {\em Phys. Rev. B} {\bf 2007}, {\em 76},~012503.
	\newblock {\url{https://doi.org/10.1103/PhysRevB.76.012503}}.
	
	\bibitem{bergeal2008pairing}
	Bergeal, N.; Lesueur, J.; Aprili, M.; Faini, G.; Contour, J.; Leridon, B.
	\newblock Pairing fluctuations in the pseudogap state of copper-oxide
	superconductors probed by the Josephson effect.
	\newblock {\em Nat. Phys.} {\bf 2008}, {\em 4},~608--611.
	\newblock {\url{https://doi.org/10.1038/nphys1017}}.
	
	\bibitem{Caprara2005}
	Caprara, S.; {Di Castro}, C.; Grilli, M.; Suppa, D.
	\newblock {Charge-fluctuation contribution to the raman response in
		superconducting cuprates}.
	\newblock {\em Phys. Rev. Lett.} {\bf 2005}, {\em 95},~9--12. 
	\href{http://xxx.lanl.gov/abs/0501671}{{\normalfont
	 [arXiv:cond-mat/0501671]}}.
	\newblock {\url{https://doi.org/10.1103/PhysRevLett.95.117004}}.
	
	\bibitem{caprara2009paraconductivity}
	Caprara, S.; Grilli, M.; Leridon, B.; Vanacken, J.
	\newblock Paraconductivity in layered cuprates behaves as if due to pairing of
	nearly free quasiparticles.
	\newblock {\em Phys. Rev. B} {\bf 2009}, {\em 79},~024506.
	
	\bibitem{leridon2020protected}
	Leridon, B.; Caprara, S.; Vanacken, J.; Moshchalkov, V.; Vignolle, B.; Porwal,
	R.; Budhani, R.; Attanasi, A.; Grilli, M.; Lorenzana, J.
	\newblock Protected superconductivity at the boundaries of charge-density-wave
	domains.
	\newblock {\em New J. Phys.} {\bf 2020}, {\em 22},~073025.
	
	\bibitem{laliberte2016origin}
	Lalibert{\'e}, F.; Tabis, W.; Badoux, S.; Vignolle, B.; Destraz, D.; Momono,
	N.; Kurosawa, T.; Yamada, K.; Takagi, H.; Doiron-Leyraud, N.;  et~al.
	\newblock Origin of the metal-to-insulator crossover in cuprate
	superconductors.
	\newblock {\em arXiv} {\bf 2016},  arXiv:1606.04491.
	\newblock {\url{https://doi.org/arXiv:1306.4583}}.
	
	\bibitem{campi2015inhomogeneity}
	Campi, G.; Bianconi, A.; Poccia, N.; Bianconi, G.; Barba, L.; Arrighetti, G.;
	Innocenti, D.; Karpinski, J.; Zhigadlo, N.D.; Kazakov, S.M.;  et~al.
	\newblock Inhomogeneity of charge-density-wave order and quenched disorder in a
	high-Tc superconductor.
	\newblock {\em Nature} {\bf 2015}, {\em 525},~359--362.
	
	\bibitem{perali1996dwave}
	Perali, A.; Castellani, C.; Di~Castro, C.; Grilli, M.
	\newblock d-wave superconductivity near charge instabilities.
	\newblock {\em Phys. Rev. B} {\bf 1996}, {\em 54},~16216--16225.
	\newblock {\url{https://doi.org/10.1103/PhysRevB.54.16216}}.
	
	\bibitem{pfleiderer2009superconducting}
	Pfleiderer, C.
	\newblock Superconducting phases of $f$-electron compounds.
	\newblock {\em Rev. Mod. Phys.} {\bf 2009}, {\em 81},~1551--1624.
	\newblock {\url{https://doi.org/10.1103/RevModPhys.81.1551}}.
	
	\bibitem{shi2014two}
	Shi, X.; Lin, P.V.; Sasagawa, T.; Dobrosavljevi{\'c}, V.; Popovi{\'c}, D.
	\newblock Two-stage magnetic-field-tuned superconductor–insulator transition
	in underdoped La$_{2-x}$Sr$_x$CuO$_4$.
	\newblock {\em Nat. Phys.} {\bf 2014}, {\em 10},~437--443.
	
	\bibitem{zhou2017spin}
	Zhou, R.; Hirata, M.; Wu, T.; Vinograd, I.; Mayaffre, H.; Kr{\"a}mer, S.;
	Reyes, A.P.; Kuhns, P.L.; Liang, R.; Hardy, W.;  et~al. {Spin susceptibility of charge-ordered YBa$_2$Cu$_3$O$_y$ across the upper critical field.} 
	\newblock {\em Proc. Natl. Acad. Sci. USA} {\bf 2017},
	{\em 114},~13148--13153.
	
	\bibitem{kavcmarvcik2018unusual}
	Ka\ifmmode \check{c}\else \v{c}\fi{}mar\ifmmode~\check{c}\else
	\v{c}\fi{}\'{\i}k, J.; Vinograd, I.; Michon, B.; Rydh, A.; Demuer, A.; Zhou,
	R.; Mayaffre, H.; Liang, R.; Hardy, W.N.; Bonn, D.A.;  et~al.
	\newblock Unusual Interplay between Superconductivity and Field-Induced Charge
	Order in ${\mathrm{YBa}}_{2}{\mathrm{Cu}}_{3}{\mathrm{O}}_{y}$.
	\newblock {\em Phys. Rev. Lett.} {\bf 2018}, {\em 121},~167002.
	\newblock {\url{https://doi.org/10.1103/PhysRevLett.121.167002}}.
	
	\bibitem{arpaia2021charge}
	Arpaia, R.; Ghiringhelli, G.
	\newblock {Charge order at high temperature in cuprate superconductors}.
	\newblock {\em J. Phys. Soc. Jpn.} {\bf 2021}, {\em
		90},~111005.
	
	\bibitem{chen1991low}
	Chen, C.; Cheong, S.; Werder, D.; Cooper, A.; Rupp~Jr, L.
	\newblock {Low temperature microstructure and phase transitions in
		La$_{2-x}$Sr$_x$CuO$_4$ and La$_{2-x}$Ba$_x$CuO$_4$}.
	\newblock {\em Phys. C Supercond.} {\bf 1991}, {\em 175},~301--309.
	
	\bibitem{tidey2022pronounced}
	Tidey, J.P.; Liu, E.P.; Lai, Y.C.; Chuang, Y.C.; Chen, W.T.; Cane, L.J.;
	Lester, C.; Petsch, A.N.; Herlihy, A.; Simonov, A.;  et~al.
	\newblock {Pronounced interplay between intrinsic phase-coexistence and
		octahedral tilt magnitude in hole-doped lanthanum cuprates}.
	\newblock {\em Sci. Rep.} {\bf 2022}, {\em 12},~14343.
	
	\bibitem{imry1975statistical}
	Imry, Y.
	\newblock On the statistical mechanics of coupled order parameters.
	\newblock {\em J. Phys. C Solid State Phys.} {\bf 1975}, {\em
		8},~567.
	
	\bibitem{lee2021multiple}
	Lee, S.; Collini, J.; Sun, S.X.L.; Mitrano, M.; Guo, X.; Eckberg, C.; Paglione,
	J.; Fradkin, E.; Abbamonte, P.
	\newblock {Multiple Charge Density Waves and Superconductivity Nucleation
		at Antiphase Domain Walls in the Nematic Pnictide
		${\mathrm{Ba}}_{1\ensuremath{-}x}{\mathrm{Sr}}_{x}{\mathrm{Ni}}_{2}{\mathrm{As}}_{2}$}.
	\newblock {\em Phys. Rev. Lett.} {\bf 2021}, {\em 127},~027602.
	\newblock {\url{https://doi.org/10.1103/PhysRevLett.127.027602}}.
	
	\bibitem{li2019tuning}
	Li, Y.; Terzic, J.; Baity, P.; Popovi{\'c}, D.; Gu, G.; Li, Q.; Tsvelik, A.;
	Tranquada, J.M.
	\newblock {Tuning from failed superconductor to failed insulator with
		magnetic field}.
	\newblock {\em Sci. Adv.} {\bf 2019}, {\em 5},~eaav7686.
	
	\bibitem{venditti2023montecarlo}
	Venditti, G.; Maccari, I.; Lorenzana, J.; Caprara, S.
	\newblock {Thermodynamic phase diagram of the competition between superconductivity and charge order in cuprates} 
	\newblock {\em In preparation}.
	
\end{thebibliography}

\end{document}